\documentstyle[aps,epsf]{revtex}

\begin{document}
\title{Vacuum decay in quantum field theory}
\author{Esteban Calzetta}
\address{Departamento de F\'{\i}sica,\\
Universidad de Buenos Aires, Ciudad Universitaria,\\
1428 Buenos Aires, Argentina}
\author{Albert Roura and Enric Verdaguer \thanks{
Also at Institut de F\'\i sica d'Altes Energies (IFAE), Barcelona, Spain.}}
\address{Departament de F\'{\i}sica Fonamental,\\
Universitat de Barcelona, Av. Diagonal 647,\\
08028 Barcelona, Spain}
\maketitle

\begin{abstract}
We study the contribution to vacuum decay in field theory due to the
interaction between the long and short-wavelength modes of the field. The
field model considered consists of a scalar field of mass $M$ with a
cubic term in the potential. The dynamics of the long-wavelength
modes becomes diffusive in this interaction. The diffusive behaviour is
described by the reduced Wigner function that characterizes the state of the
long-wavelength modes. This function is obtained from the whole Wigner
function by integration of the degrees of freedom of the short-wavelength
modes. The dynamical equation for the reduced Wigner function becomes a
kind of Fokker-Planck equation which is solved with suitable boundary
conditions enforcing an initial metastable vacuum state trapped in the
potential well. As a result a finite activation rate is found, even at zero
temperature, for the formation of true vacuum bubbles of size $M^{-1}$. This
effect makes a substantial contribution to the total decay rate.
\end{abstract}

\section{Introduction}

\label{i}

In this paper we report our preliminary findings within a larger program
which aims at the development of a theory of non equilibrium first order
phase transitions, such as have occurred in the Early Universe (grand
unified and electroweak symmetry breaking \cite{GHS}) and, possibly, in the
first stages of heavy ion collisions (chiral symmetry breaking and
confinement \cite{Shuryak,gleiser}). For this reason, we must seek a
description of the decay process which emphasizes the dynamical aspects,
over the static aspects encoded in the effective potential.

Vacuum decay in field theory is described with a potential which displays a
local minimum, separated from the absolute minimum by a potential barrier. A
system prepared in the false vacuum state (metastable phase) within the
potential well may decay in essentially two different ways, namely (a) by
tunneling effect, that is, by going through the barrier in a classically
forbidden trajectory \cite{Gurney,Landau,Migdal}, or else, (b) by
activation, that is, by jumping above the barrier \cite{Kramers,Langer1}. In
either case, the decay probability follows the law $P\sim A\exp(-B)$ which
gives the probability per unit time and unit volume to nucleate a region of
the stable phase within the metastable phase. In the tunneling effect,
$B=S/\hbar$, where $\hbar$ is Planck's
constant and $S$ is the action for the
trajectory which goes under the barrier in imaginary time \cite{Coleman}. In
thermal activation, $B=V_s/(k_B T)$, where $k_B$ is Boltzman's constant, $T$
the temperature, and $V_s$ is the height of the free energy measured from
the false vacuum \cite{HTB,Affleck,Jaume}. Thus, activation disappears as $
T\to 0$.

In systems with few degrees of freedom, there must be an external agent,
typically a thermal source, for activation to be possible. Our thesis is
that in field theories there is a phenomenon similar to activation even at
zero temperature. This comes from the observation that in a field theory,
when a mode decomposition of the field is made, there are only a few long
wavelenghts modes which are unstable and decay. These nearly homogeneous
modes may be regarded as an open system, which interacts with the
environment provided by the shorter wavelength modes. It is then possible to
describe the quantum evolution of the system in terms of an effective
dynamics, whereby the interaction with the environment results in the onset
of dissipation and noise.

The ultimate reason for the presence of a finite activation rate even at
zero temperature is that, for a generic field theory the dynamics of these
homogeneous modes is anharmonic enough to contain Fourier components with
frequencies above the threshold for excitation of the short wavelength
modes. This results in an energy transfer from the long wavelength or
homogeneous modes to the short wavelength or inhomogeneous modes through
particle creation. As demanded by the energy balance, and encoded in the
fluctuation-dissipation theorem, this energy flow is compensated by a
stochastic force on the homogeneous modes, originated in fluctuations of the
inhomogeneous modes. Thus the dynamics of the homogeneous modes becomes
diffussive, even if, properly speaking, there is no external ``environment''
to the field \cite{CH94}.

We wish to stress that this is not only a theoretical possibility. In this
paper, we will show through a detailed analysis of a concrete model that the
activation rate makes a substantial contribution to the full decay rate even
at zero temperature. In the process, we shall develop the necessary
formalism to compute the activation rate to leading order in $\hbar$.

The key ingredient will be the description of the quantum state of the long
wavelength modes of the field by means of the reduced Wigner function. This
function has the same information that the reduced density matrix of an open
quantum system but is similar in many aspects to a distribution function in
phase-space. The dynamical equation for the reduced Wigner function (master
equation) includes noise terms produced by the short wavelength modes to
quadratic order in the parameter coupling the short and long wavelength
modes. We shall derive the tunneling rate from an analysis of the decay of
non trivial solutions of the master equation, after suitable boundary
conditions have been enforced.

The master equation contains all required information on vacuum decay,
including both the ``tunneling" aspect of the homogeneous mode as well as
``activation", {\it i.e.} the effect due to the back reaction of the
inhomogeneous modes, but in this paper we concentrate in the activation
aspect. Thus, whereas the instanton method \cite{Coleman}, for instance,
seems to be best suited to describe the pure tunneling of the homogeneous
mode, the effect due to the back reaction of the inhomogeneous modes seems
to be best described by the present method. Another advantage of this
approach, which was one of our first motivatons, is that whereas the
instanton method works well for processes not far from equilibrium, this
method should work also for out of equilibrium situations since it is based
on a real time evolution equation such as the master equation.

Our conclusion shall be that, for a physically well motivated
system-environment separation, the contribution to the exponent coming from
quantum fluctuations in the short wavelength modes is comparable to, but
distinguishable from, the contribution from tunneling under the barrier by
the long wavelength modes themselves. Which contribution is actually
dominant will depend on the specifics of each model. Therefore, the
approaches that underplay backreaction from the short wavelenght modes,
underestimate the tunneling rate, and may even miss the largest contribution.

Comparing with the literature on tunneling in open and driven systems, the
main difference is that we do not assume a priori any features of the noise
and dissipation, but rather derive them from the underlying unitary field
theory. In practice, this means that we shall have to deal with non-local
dissipation and coloured noise. Also for simplicity, we shall focus on
computing the exponent in the tunneling rate to leading order in $\hbar$,
and at zero temperature. As a matter of fact, this is the difficult case,
since at high temperature the interaction of the short and long wavelength
modes is just another aspect of the interaction with the heat bath.

The paper is organized as follows. In Sec. \ref{i1} we briefly review
current theories of vacuum decay and place our work in that context. In Sec.
\ref{ii} we present the basic derivation of the master equation for the
reduced Wigner function for a Quantum Brownian Motion (QBM) model which is
of relevance for our problem. The model is an open quantum system consisting
of of a massive particle in an arbitrary potential coupled to an infinite
set of harmonic oscillators, the coupling is linear in the system variables
but quadratic in the oscillator coordinates. For our problem the master
equation reduces to a Fokker-Planck like equation. In Sec. \ref{iii} we
consider a field theory model consisting of a massive scalar field with a
negative cubic potential term. The long and short wavelength modes of the
field are separated and we reduce the problem to a system-environment
interaction similar to the QBM model discussed in the previous section. The
relevant kernels for the corresponding Fokker-Planck equation are
calculated. Sec. \ref{iv} contains the derivation of the tunneling rate from
the analysis of the solutions to the Fokker-Planck equation. We discuss our
results in the final section. A number of appendices carry some of the
unavoidable technical details.

\section{Vacuum decay: a survey}

\label{i1}

\subsubsection{Vacuum decay in systems with one degree of freedom}

The twin issues of thermal activation and spontaneous nucleation have a long
and distinguished history; see Refs. \cite{GSMS,HTB} for a review. In its
simplest formulation, we deal with a quantum mechanical system with one
degree of freedom $x$ and Hamiltonian $H=p^2/2+V( x) $, where $V$ has the
generic form shown in the figure. It is an experimental fact that even if we
prepare the system to be confined to the neighborhood of the ``false
vacuum'' $x\sim 0$, and the barrier is much higher than the typical energies
accesible to the system ($E\sim \hbar V^{\prime \prime }\left( 0\right) $ in
the quantum mechanical problem, $E\sim kT$ at finite temperature), the
system will find a way to escape from the potential well after some typical
time $\tau $ has ellapsed. The problem is to estimate the ``mean life''
$\tau $ or equivalently the ``tunneling rate'' $\tau ^{-1}$.

Vacuum decay can be formulated as a steady state problem if we inject
particles into the system in order to keep a constant population in the
false vacuum state. We then have a constant flux of particles impinging on
the barrier from the left, and the problem reduces to the computation of the
transmission coefficient \cite{Gurney}. In quantum mechanics, this is
readily obtained within the WKB approximation, and the result is the
Arrhenius-like expression
\begin{equation}
\tau ^{-1}\sim \Delta \exp (-S/\hbar), \qquad S=\int_0^{x_{exit}}dx\;
\sqrt{2V\left( x\right) },  \label{arrhenius}
\end{equation}
where $\Delta $ is a prefactor of order $1$, and we have set the classical
energy of the false vacuum to $0$.

This formula describes vacuum decay through tunneling, that is, an
essentially quantum phenomenon. If we allow the system to interact with an
external noise source (typically a heat bath at a given temperature $T$),
then the energy of the system alone is no longer conserved, and the system
can jump over the barrier, resulting in vacuum decay through activation. The
activation rate has been computed, within the ``constant flux'' approach, by
Kramers \cite{Kramers} and Langer \cite{Langer1}. They show that equation
(\ref{arrhenius}) still holds, but the exponent $S/\hbar $ becomes $F_s/kT$,
where $F_s$ is the activation free energy, {\it i.e.} the height of the free
energy barrier to be overcome.

A less artificial approach attempts to compute the actual time evolution of
the false vacuum state $\Psi _F$. This is not a stationary state of the
system, but it may be expanded in energy eigenstates $\psi _E$. The
amplitude of $\psi _E$ in the expansion of $\Psi _F$ peaks around $E\sim 0$,
and for small energy may be approximated by the Breit-Wigner form \cite
{Landau,Migdal}
\begin{equation}
\Psi _F\left( x,t\right) \sim \int dE\;e^{-iEt/\hbar }\frac{\psi _E\left(
x\right) }{E^2+\lambda ^2}.  \label{breit}
\end{equation}
At long, but not too long, times we obtain $\Psi _F\sim Ce^{-\lambda t/\hbar
}$. The false vacuum behaves as an energy eigenstate with complex energy
$E_F\sim -i\lambda $ (complex energies and eigenstates may be defined within
an extended formalism of quantum mechanics \cite{Mario}), and the mean life
is $\tau \sim \hbar \lambda ^{-1}$.

The calculation of activation rates in the ``complex energy'' approach has
been worked out by Langer \cite{Langer2}. The idea is to define a ``free
energy'' for an ensemble of unstable configurations including the critical
droplet. This ``free energy'' is complex, and its imaginary part is related
to the mean life as in the quantum mechanical problem. The physical free
energy, of course, is real, and it is given by Maxwell's construction \cite
{Maxwell}.

Coleman and collaborators have proposed a simple and elegant way to compute
the complex false vacuum energy \cite{Coleman}. The idea is that the vacuum
energy can be expressed in terms of a path integral over Euclidean histories
with appropiate boundary conditions. For unstable systems, the path integral
must be computed by analytic continuation, and an imaginary part appears. In
certain cases it is possible to show that the path integral is dominated by
the contributions from a discrete set of saddle points, corresponding to
sequences of ``bounces'' against the inner sides of the barrier with little
or no overlap between bounces. Then the formula in Eq. (\ref{arrhenius}) is
recovered, where $S$ is now interpreted as the euclidean action for the one
bounce solution, also called the ``instanton''.

The instanton method is easily generalized to the thermal case
\cite{Affleck}. The idea is to write the partition function
for the unstable system as a
path integral over Euclidean configurations with periodicity $\beta \hbar $
($\beta =1/kT$) in Euclidean time, and then to evaluate the path integral in
the saddle point approximation. Due to new boundary conditions, the thermal
instanton may not be the same as the instanton at $T=0$. The change in the
nature of the instanton gives a simple and compelling interpretation of the
crossover from spontaneous transition to thermal activation \cite{Jaume}.

The tunneling rate can also be derived from the large order behavior of
perturbation theory \cite{suzuki}

\subsubsection{Tunneling in systems with few degrees of freedom}

All approaches discussed so far have natural generalizations to systems with
few degrees of freedom. In the case of the instanton approach, the
generalization is almost immediate, only one has to take care of symmetries
of the system which may appear as zero modes in the spectrum of
perturbations around the instanton solution, thus causing an apparent
divergence of the path integral. These symmetries may be handled by
isolating them as collective modes prior to the saddle point evaluation of
the path integral \cite{Bes}.

The ``constant flux'' approach is implemented by seeking a solution to the
Schr\"odinger equation within the WKB or Born-Oppenheimer approximation \cite
{Banks}. The idea is to identify a single variable $x$ which parametrizes
the ``most probable scape path'', namely the path across the saddle
separating the false and true vacua. Then one uses a mixed ansatz for the
wave function, whereby it is assumed to be Gaussian on all other variables,
and of WKB form with respect to $x$. ``Under the barrier'', the WKB\
approximation leads to a Hamilton-Jacobi equation for a particle moving in
an inverted potential, the same dynamical problem one confronts in the
instanton method, although now the potential may be modified by terms
arising from the zero point energies of the transversal modes; but not of $x$
itself \cite{Gleiser}. This solution is then matched to oscillatory
solutions on either side of the barrier.

Conceptually, the WKB method has some advantages over the instanton method
\cite{Goncharov}. In the first place, the connection to the Schr\"odinger
equation is much more straightforward. Also, one has more control on the
quantum state of the transversal degrees of freedom, which then allows one
to ask questions like whether tunneling is associated to particle creation
\cite{Rubakov,Tanaka,Laura}. Finally, it has proved easier to define the
range of validity of the approximations involved in the WKB framework than
in the Euclidean path integral one. However, they are likely to be
equivalent in last analysis, and the conclusion that these methods work best
when the tunneling variable $x$ is slow compared with the transversal modes
probably holds equally well for both approaches \cite{Schmid}. We refer the
reader to Ref. \cite{Bonini} for recents developments along these lines, and
to Ref. \cite{JPS} for an example of similar techniques in a different
context.

Some nonequilibrium aspects of vacuum decay have been the focus of work by
Boyanovsky et al. \cite{dan}. They show that realistic initial conditions
usually imply a nonzero probability for the system to be at the unstable
configuration in the saddle of the free energy surface. Starting from this
configuration, the roll down of the system towards the true vacuum may be
analyzed by usual nonequilibrium field theory methods. However, the initial
amplitude is still computed by conventional methods, such as discussed so
far.

\subsubsection{Tunneling in quantum field theory}

Gervais, Sakita and De Vega have applied the WKB method to tunneling in
quantum field theory \cite{Gervais}; see Refs. \cite{Affleck2,Weiss} for
instanton and complex-time methods. In spite of the obvious similitudes,
there are some important differences between the problem of tunneling in
field theory and in systems with few degrees of freedom. Some of these
differences are technical in nature, such as the need to carefully account
for loop corrections to the effective potential \cite{Buchmuller}, and to
adopt a regularization procedure to compute the prefactor in the Arrhenius
formula \cite{Kristin,Munster,Kiselev}, which would be divergent if naively
computed. There is also a fundamental conceptual difference which we now
discuss.

A field theory only makes sense, from the physical point of view, if it is
understood as an effective theory describing the low frequency sector of a
more fundamental theory, whose high frequency degrees of freedom may be
totally unlike continuous fields; such as strings in elementary particle
physics, discrete lattices in condensed matter applications, molecules in
hydrodynamics \cite{Weinberg}.

One clear way to bring this point home is by explicitly integrating out all
modes with wave number $k>\Lambda $, where $\Lambda $ is some cutoff. The
remaining modes are described by a coarse grained effective potential \cite
{Wegner,Wetterich,Strumia,Attanasio,Liao,Diego}. As $\Lambda $ decreases
from $\infty $ to $0$, the coarse grained effective potential interpolates
between the ``bare'' potential shown in the figure to the Maxwell construction
free energy \cite{Aoki}, showing the drastic effect of the short wavelength
modes or high frequency sector on the physics of the long wavelength modes
or low frequency sector. At some point the barrier separating the different
metastable points dissapears, reflecting the effect of averaging the field
over distances much larger than a domain.

\subsubsection{Tunneling in open systems}

As we have seen from earlier studies of the coarse grained effective
potential, tunneling in field theory should be properly posed as an open
quantum system problem. However, the nature of tunneling in an open system
cannot be described adequately by just computing changes in an effective
potential. Besides the static changes reflected by the scale dependent
effective potential, the dynamics of the long wavelength modes will become
both dissipative and stochastic. The onset of dissipation and noise is also
generic to semiclassical \cite{CH94} and effective theories \cite{CH97}.

For the present discussion it is essential to realize that noise and
dissipation are actually two aspects of a single phenomenon, the dynamical
action and back-reaction between ``system'' and ``environment''. In
equilibrium situations, this inner relationship can be made explicit through
the ``fluctuation-dissipation" relation \cite{FDT}.

A simple way to deal with tunneling in open systems is to model the
environment explicitly within a larger system-environment complex, in effect
reducing the problem to tunneling in many degrees of freedom \cite
{Caldeira,Weiss2,Bray}. However, it is essential to avoid approximations
(such as assuming that the environment degrees of freedom perform linear
oscillations around a prescribed trajectory of the system) which in practice
underplay the back reaction of the environment on the system, and thus break
the balance between fluctuation and dissipation. The relevance of the
fluctuation-dissipation relation to tunneling has been emphasized in refs.
\cite{fujikawa}.

The broadening of the reduced Wigner function of the open system by external
noise has been discussed in Ref. \cite{matsumoto}. Unlike the present work,
these authors consider an external noise source, whose spectral features may
be chosen at will. In order to give an adequate account of back reaction,
Bruinsma and Bak \cite{Bak} have proposed treating the system as propagating
in a random medium, the randomness being associated to the environmental
variables. In a second step, a path integration over histories of the bath
allows the computation of the tunneling rate. As in the present work, a
serious consideration of back reaction leads to describing the system as a
driven system, subject to stochastic forces originating from the environment.

The theory of vacuum decay in open systems has points of contact with the
problem of decay in driven systems \cite{Horsthemke,Grifoni}, although in
these later studies usually the properties of the driving force are assumed
a priori, rather than derived from a more comprehensive model. It is also
possible to obtain a path integral representation of the solution of a
Langevin equation, whereby an open system may be subsumed into a larger
field theory \cite{Zinn-Justin,Carmen}. See Ref. \cite{Gabo} for further
developments.

\subsubsection{Tunneling and the semiclassical approximation}

Semiclassical field theories, where some fields are treated as c-number,
while the rest is described quantum mechanically, may be seen as open
systems, with the classical fields as the system and the quantum fields as
the environment. Quantum fluctuations in the system are registered as noise
by the environment, and may induce transitions.

An early application of these ideas appeared within Starobinsky's
``stochastic inflation'' program \cite{Starobinsky}. Here the superhorizon
modes of the inflaton field during inflation are the system whereas all
other shorter wavelength modes are the environment; for a discussion of the
validity of the semiclassical approximation, see \cite{CH95,Kiefer}.
Cosmological redshift causes a continuous streaming of modes from the
environment to the system, which may be regarded as a white noise source.
This noise may allow the system to hop over potential barriers, seeking the
absolute minima. Eventually, a steady state distribution of cosmological
domains is reached, not unlike that predicted by the Hartle and Hawking
``wave function of the Universe'' \cite{Goncharov,Mijic}.

Much more generally, within the semiclassical approximation the back
reaction to cosmological particle creation processes always has a stochastic
component \cite{CH94}, for which reason the correct semiclassical
description of the Early Universe ought to be formulated in terms of an
stochastic ``Einstein-Langevin'' equation \cite{ELE}. If we consider an
ensemble of Universes, then we may introduce a distribution function obeying
a Fokker-Planck-like equation \cite{CV}. This equation describes activation
phenomena, which are the semiclassical version of Vilenkin's ``creation from
nothing'' scenario \cite{Vilenkin}.

\section{Open systems and the reduced Wigner function}

\label{ii}

Before dealing with field theory we will consider in this section a Quantum
Brownian Motion (QBM) model which is typically used as a a paradigm of an
open quantum system. The system has an arbitrary potential and the coupling
between the system and the environment is linear in the system variables but
quadratic in the environment variables. That feature will be of relevance
when dealing in section \ref{iii} with our field-theory model. The main
result of this section is the derivation of the master equation for the
reduced Wigner function of the QBM model to leading order in $\hbar$. This
equation turns out to be a Fokker-Planck equation which is similar to that
used by Kramers \cite{Kramers} to study the activation problem in
statistical physics.

\subsection{A QBM model}

As our QBM model we consider a system consisting of a particle of unit mass
($M=1$) described with a variable $x$ and subject to an arbitrary potential
with a quadratic part corresponding to an oscillator of frequency $\Omega_0$
and an anharmonic part $V^{(nl)}(x)$, {\it i.e.} $V(x)=(1/2)\Omega_0^2\; x^2
+V^{(nl)}(x)$, which is coupled to an environment consisting of an infinite
set of harmonic oscillators with coordinates $q_j$. The action for the whole
set of degrees of freedom is defined by
\begin{equation}
S[x,\{q_j\}]= S[x]+S[\{q_j\}]+S_{int}[x,\{q_j\}],  \label{1.1}
\end{equation}
where the system, environment and interaction actions are given,
respectively, by
\begin{eqnarray}  \label{1.3}
S[x]&=& \int dt\left({\frac{1}{2}} \dot x^2- V(x)\right), \\
S[\{q_j\}]&=& \sum_j\int dt\left({\frac{1}{2}} m\dot q_j^2-{\frac{1}{2}}
m\omega_j^2q_j^2\right), \\
S_{int}[x,\{q_j\}]&=&\sum_j g\int dt\;xq_j^2,  \label{1.4}
\end{eqnarray}
where $g$ is a coupling constant, and we have assumed that the coupling is
linear in the system variable but quadratic in the environment variables.
The environment oscillators have all the same mass $m$ and their frequencies
are $\omega_j$. At this point the potential $V^{(nl)}(x)$ is arbitrary, but
later we will take a cubic potential, $V^{(nl)}(x)=-(\lambda/6)x^3$, in this
way the total potential will present a local minimum and a barrier as
required to represent the system in a metaestable phase. Also the parameters
$\lambda$ and $g$ are unrelated, however when we consider the
application to a field theory these parameters will coincide.

The reduced density matrix for our open quantum system at a certain final
time $t_f$ is defined from the density matrix $\rho$ of the whole system by
tracing out the environment degress of freedom at that time
\begin{equation}
\rho_r(x_f,x_f^{\prime}, t_f)=\int\prod_j dq_j\rho(x_f, \{q_j\},
x_f^{\prime},\{q_j\},t_f).  \label{1.5}
\end{equation}

The reduced density matrix at time $t_f$ can be written in terms of the
reduced density matrix at the initial time $t_i$ by the evolution equation,
\begin{equation}
\rho_r(x_f,x_f^{\prime}, t_f)=\int dx_i dx_i^{\prime}J(x_f,x_f^{\prime},
t_f;x_i,x_i^{\prime}, t_i)\rho_r(x_i, x_i^{\prime},t_i),  \label{1.6}
\end{equation}
in terms of the propagator $J$, whose path integral representation is
\begin{equation}
J(x_f,x_f^{\prime}, t_f;x_i,x_i^{\prime}, t_i)= \int_{x_i}^{x_f} {\cal D}
x\int_{x_i^{\prime}}^{x_f^{\prime}} {\cal D}x^{\prime}
\exp {\frac{i}{\hbar}}
\left( S[x]-S[x^{\prime}] + S_{IF}[x,x^{\prime}]\right),  \label{1.7}
\end{equation}
where $x_i=x(t_i)$, $x_f=x(t_f)$ and similarly for the primed quantities,
and $S_{IF}[x,x^{\prime}]$ is the Feynman and Vernon influence action \cite
{feynman}. When the system and the environment are initially uncorrelated
the initial density matrix factorizes, {\it i.e.} $\rho(t_i)=\rho_r(t_i)
\rho_e(t_i)$ (here $\rho_e$ stands for the environment density matrix), the
influence functional, which is defined by $F[x,x^{\prime}]=\exp
(iS_{IF}[x,x^{\prime}]/\hbar)$, can be expressed by
\begin{eqnarray}
F[x,x^{\prime}]&=&\prod_j\int dq_j^{(f)} dq_j^{(i)}dq_j^{\prime(i)}
\int_{q_j^{(i)}}^{q_j^{(f)}} {\cal D} q_j \int_{q_j^{\prime(i)}}^{q_j^{(f)}}
{\cal D} q_j^\prime \exp {\frac{i}{\hbar}}\left(
S[\{q_j\}]-S[\{q_j^\prime\}] +
S_{int}[x,\{q_j\}]-S_{int}[x^\prime,\{q_j^\prime\}]\right)  \nonumber \\
&&\cdot\rho_e(\{q_j^{(i)}\},\{q_j^{\prime(i)}\},t_i),  \label{1.8}
\end{eqnarray}
where $q_j^{(i)}=q_j(t_i)$, $q_j^{\prime(i)}=q_j^\prime(t_i)$, and at the
final times $q_j(t_f)=q_j^{(f)}=q_j^\prime(t_f)$.

Assuming a Gaussian
initial state for the environment, $\rho_e$ is Gaussian and the influence
functional can be computed perturbatively in $g$ from the path integral. Up
to second order in $g$ \cite{feynman} we have for the influence action,
\begin{equation}
S_{IF}[x,x^\prime]= -2\int_{t_i}^{t_f}dt\int_{t_i}^t dt^\prime \Delta(t)
D(t,t^\prime)X(t^\prime)+{\frac{i}{2}} \int_{t_i}^{t_f}dt\int_{t_i}^{t_f}
dt^\prime \Delta(t) N(t,t^\prime)\Delta(t^\prime),  \label{1.9}
\end{equation}
where we have introduced the average and difference coordinates defined,
respectively, by
\begin{equation}
X(t)\equiv {\frac{1}{2}}(x^\prime(t)+x(t)),\ \ \ \Delta(t)\equiv x^\prime
(t)-x(t).  \label{1.9a}
\end{equation}
The kernels $D(t,t^\prime)$ and $N(t,t^\prime)$ are called dissipation and
noise kernels, respectively, and are defined by $D=\sum_j D_j$ and $N=\sum_j
N_j$ where
\begin{eqnarray}
D_j(t,t^\prime)&=& -{\frac{i}{2}}\langle \left[
\Xi_j(t),\Xi_j(t^\prime)\right]\rangle,  \label{1.10} \\
N_j(t,t^\prime)&=&{\frac{1}{2}}\langle \left\{
\Xi_j(t),\Xi_j(t^\prime)\right\}\rangle -\langle\Xi_j(t)\rangle
\langle\Xi_j(t^\prime)\rangle,  \label{1.11}
\end{eqnarray}
with $\Xi_j= gq_j^2$. It is now convenient to introduce the kernels $
H_j(t,t^\prime)=-2D_j(t,t^\prime) \theta(t-t^\prime)$ and we can write the
influence action in the form
\begin{equation}
S_{IF}[x,x^\prime]= \Delta\cdot H\cdot X +{\frac{i}{2}}\Delta\cdot N\cdot
\Delta,  \label{1.12}
\end{equation}
where we have introduced the notation $A\cdot B\equiv \int dt A(t)B(t)$ and
defined $H=\sum_j H_j$ which we may write formally as $H(t,t^
\prime)=-2D(t,t^\prime) \theta(t-t^\prime)$. This last equality is however a
formal expression since being the product of two distributions, $H$ is not
well defined and suitable regularization and renormalization is required.
This term, in fact, has local divergent parts that may be reabsorbed into
the parameters of the bare action, see \cite{RV99} for details. Thus, from
now on we will assume that $H$ is a well defined distribution in the
previous sense.

\subsection{The reduced Wigner function}

Our main purpose in this subsection is to write the reduced Wigner function
for the system in a suitable way. This is a phase space function which is
defined from the reduced density matrix by the following integral transform
\begin{equation}
f(X,p,t)={\frac{1}{2\pi \hbar }}\int_{-\infty }^{\infty }d\Delta e^{ip\Delta
/\hbar }\rho _{r}(X-\Delta /2,X+\Delta /2,t).  \label{1.13}
\end{equation}

The reduced density matrix (\ref{1.6}) at time $t_f$ can be computed from
the path integrals of (\ref{1.7}). To carry out this computation we will
follow closely Ref. \cite{CRV00} where a similar computation for a linear
system coupled linearly to an environment was described. For this reason we
will describe here only the main steps and will concentrate on those which
are peculiar to the nonlinear system. This is performed in several steps in
which a key role is played by the use of the coordinates $X$ and $\Delta$
instead of $x$ and $x^\prime$. The first step is to integrate the system
action in (\ref{1.7}) by parts using the new coordinates $X$ and $\Delta$,
\begin{equation}
S[x]-S[x^{\prime }]=-\dot{X}\Delta
|_{t_{i}}^{t_{f}}+\int_{t_{i}}^{t_{f}}dt\;\Delta (t)\left( {\frac{d^{2}}
{dt^{2}}}X(t)+\left. {\frac{\partial V}{\partial X}}\right| _{\Delta
=0}\right) +\dots ,  \label{1.13a}
\end{equation}
where the ellipsis stands for the terms nonlinear in $\Delta$ that come from
the potential $V(x)$ which involve higher derivatives in $X$ (see below).
Note that due to the fact that the potential gradient
is evaluated at $\Delta=0$ this term can also
be written as $V^\prime(X)$ where $V(X)$ is functionaly the same as $V(x)$.

The change of integration variables $\int_{x_i}^{x_f}{\cal D}x
\int_{x_i^\prime}^{x_f^\prime}{\cal D}x^\prime$ $\to$ $\int_{X_i}^{X_f}{\cal
D}X \int_{\Delta_i}^{\Delta_f}{\cal D}\Delta$ involves a Jacobian which is
unity and thus the path integration of the propagator (\ref{1.7}) can be
written as
\begin{equation}
\int_{X_{i}}^{X_{f}}{\cal D}X\int_{\Delta _{i}}^{\Delta _{f}}{\cal D}\Delta
e^{\frac{i}{\hbar }\Delta \cdot L[X]}F[\Delta ,X],  \label{1.14}
\end{equation}
where $L[X]$ is a functional of $X$ and a function of $t$ defined by
\begin{equation}
L[X;t)\equiv \left( {\frac{d^{2}}{dt^{2}}}X(t)+
\left. {\frac{\partial V}{\partial X}}\right| _{\Delta =0}\right)
+\int_{t_i}^t dt^{\prime }H(t,t^{\prime})X(t^{\prime }),
\label{1.14a}
\end{equation}
and the functional $F[\Delta ,X]$ incorporates in the exponent all the terms
that are not linear in $\Delta $ which come from the influence action and
from the nonlinear potential of the system action, when expressed in
the variables $X(t)$ and $\Delta (t) $. More precisely,
\begin{equation}
F[\Delta ,X]=\exp \frac{i}{\hbar }\left( {\frac{i}{2}}\Delta \cdot N\cdot
\Delta +\int_{t_i}^{t}dt^{\prime }\;V^{(nl)}[\Delta ,X]\right) \text{,}
\label{1.15}
\end{equation}
where
\begin{equation}
V^{(nl)}[\Delta ,X]=2\sum_{n\geq 1}\frac{1}{(2n+1)!}\left. \frac{\partial
^{(2n+1)}V}{\partial X^{(2n+1)}}\right| _{X,\Delta =0}(-\Delta /2)^{2n+1}
\text{.}  \label{1.16}
\end{equation}
In particular, for the cubic potential $V^{(nl)}(x)=-(\lambda /6)x^{3}$ we
have $V^{(nl)}[\Delta ,X]=-(\lambda /24)\Delta ^{3}$.

Let us now introduce
the functional Fourier transform
\begin{equation}
P[\xi ]=K\int {\cal D}\Delta \;e^{i\Delta \cdot \xi /\hbar }e^{-\Delta \cdot
N\cdot \Delta /2\hbar }\text{,}  \label{1.17}
\end{equation}
where $K=1/\det (2\pi \hbar I)$, the interpretation of this functional will
be discussed below. With expression (\ref{1.17}) we may write the reduced
Wigner function as
\begin{eqnarray*}
f(X_{f},p_{f},t_{f}) &=&{\frac{1}{2\pi \hbar }}\int dX_i d\Delta_i
\int {\cal D}\xi \;P\left[
\xi \right] \;\int_{-\infty }^{\infty }d\Delta _{f}\;e^{ip_{f}\Delta
_{f}/\hbar }\int_{X_{i}}^{X_{f}}{\cal D}X\;e^{-i\Delta _{f}\dot{X}\left(
t_{f}\right) /\hbar } \\
&&\int_{\Delta _{i}}^{\Delta _{f}}{\cal D}\Delta \;e^{i\Delta
\cdot \left( L[X]-\xi \right)/\hbar }
e^{i\int V^{(nl)}/\hbar }e^{i\Delta _{i}
\dot{X}\left( t_{i}\right) /\hbar }\rho _{r}(X_{i}-\Delta
_{i}/2,X_{i}+\Delta _{i}/2,t_{i});
\end{eqnarray*}
and using that
$
\Delta =\exp ( i \Delta \cdot \xi /\hbar)
\left( i\hbar  \delta /\delta\xi\right)
\exp (-i \Delta \cdot \xi/\hbar ),
$
it may be rewritten as
\begin{eqnarray*}
f(X_{f},p_{f},t_{f}) &=&{\frac{1}{2\pi \hbar }}\int dX_i d\Delta_i
\int_{X_{i}}^{X_{f}}{\cal D}
X\int {\cal D}\xi \;P_{Q}[ \xi ,X;t_{f}) \;\int_{-\infty
}^{\infty }d\Delta _{f}\;e^{i\left( p_{f}-\dot{X}\left( t_{f}\right) \right)
\Delta _{f}/\hbar } \\
&&\ \int_{\Delta _{i}}^{\Delta _{f}}{\cal D}\Delta \;
e^{i\Delta \cdot \left( L[X]-\xi \right)/\hbar }
e^{i\Delta _{i}\dot{X}\left(
t_{i}\right) /\hbar }\rho _{r}(X_{i}
-\Delta _{i}/2,X_{i}+\Delta _{i}/2,t_{i}),
\end{eqnarray*}
where
\begin{equation}
P_{Q}[ \xi ,X;t) =\left\{ \exp \left[ \frac{i}{\hbar }
\int_{t_i}^{t}dt^{\prime }\;
V^{(nl)}[-i\hbar \frac{\delta }{\delta \xi },X]\right]
\right\} P\left[ \xi \right]\text{.}
\label{1.17a}
\end{equation}

A convenient way to perform the path integration for $X(t)$
is to introduce the following functional change:
\begin{equation}
X(t)\to \left\{ X_i=X(t_i), p_i= \dot X(t_i), \tilde \xi (t)=
L[X;t)\right\}.  \label{1.18}
\end{equation}
With this transformation the function $X(t)$ becomes substituted by the
initial conditions ($X_i,p_i$) and the function $\tilde \xi(t)$ in the path
integration. This functional change is invertible, in the sense that
$\{X_i,p_i,\tilde\xi(t)\}$ $\to$ $X(t)$, since the solution $X(t)$ of the
integro-differential equation involved in (\ref{1.18}) is unique given
initial conditions ($X_i,p_i$).
A subtler point concerns the Jacobian of the transformation (\ref{1.18}).
Even though this transformation is nonlinear one can show that the Jacobian
is constant. This can be seen by discretizing the time $t_{k}=\epsilon
k+t_{i}$ ($k=1,2,\dots ,n$, and $t_{i}$ is the initial time). Then the
corresponding values ($X_{i},X_{1},\dots ,X_{n}$), where $X_{k}=X(t_{k})$,
map into ($X_{i},p_{i},\tilde{\xi}_{2},\dots ,\tilde{\xi}_{n}$) in such a
way that the Jacobian matrix has zero elements above the diagonal. For
instance, the second derivative terms become $
[(X_{k}-X_{k-1})-(X_{k-1}-X_{k-2})]/\epsilon ^{2}+V^{\prime
}(X_{k-1})-\sum_{k^{\prime }<k}H_{kk^{\prime }}X_{k^{\prime }}=
\tilde\xi _{k}$.
The Jacobian is thus the product of the diagonal elements, which are
constant (independent of any $X_{k}$). Then one may write $\int {\cal D}
X\dots =\bar{K}\int dX_{i}dp_{i}\int {\cal D}\tilde{\xi}\dots $ and
introduce convenient delta functions such as $\delta (X(t_{f})-X_{f})$ to
ensure that the correct final points appearing in (\ref{1.14}), {\it i.e.},
$\int^{X_{f}}{\cal D}X$ are recovered from the functional integral $\int
{\cal D}\tilde{\xi}$ with free ends. One should also be careful about the
dependence on the initial conditions $(X_{i},p_{i})$ in the general case.

Now we first perform the integral ${\cal D}\Delta$ which simply leads
to a term proportional to $\delta(\tilde\xi-\xi)$ and the integral
${\cal D}\tilde\xi$ is then trivial. On the other hand the integral
$d\Delta_i$ brings back the reduced Wigner function at the initial
time according to Eq. (\ref{1.13}). Finally, we get
the following suggestive form for the reduced Wigner function
at the final time
\begin{equation}
f(X_{f},p_{f},t_{f})=\bar{K}\int dX_{i}dp_{i}\int {\cal D}\xi \;P_{Q}
[\xi ;t_{f}) \;\delta \left( p_{f}-\dot{X}\left( t_{f}\right) \right)
\delta \left( X_{f}-X\left( t_{f}\right) \right) f(X_{i},p_{i},t_{i}),
\label{wigner}
\end{equation}
where $X(t)$ is a solution of the integro-differential equation
\begin{equation}
L[X;t)=\xi (t),  \label{langevin}
\end{equation}
with initial conditions $\left( X_i,p_i\right) $,
{\it i.e.} $X=X[\xi;X_i,p_i)$, and
\begin{equation}
P_{Q}[\xi ;X_{i},p_{i},t) =P_{Q}[ \xi ,X [\xi ;X_{i},p_{i});t).
\label{1.18a}
\end{equation}

The constant $\bar{K}$ form the Jacobian can be determined from the
condition that ${\rm Tr}\rho _{r}(t_{f})=1$, {\it i.e.} that $\int_{-\infty
}^{\infty }dX_{f}\rho _{r}(X_{f},X_{f},t_{f})=1$,
which is equivalent to $\int_{-\infty}^\infty dX_f dp_f
f(X_f,p_f,t_f)=1$. Inserting expression (\ref{wigner}) for
$f(X_f,p_f,t_f)$ we get
\begin{equation}
\bar K \int dX_i dp_i\int {\cal D} P_Q[\xi;t_f)
f(X_i,p_i,t_i)=1.
\label{1.18b}
\end{equation}
When $P_Q[\xi,X;t)$ does not depend on $X$, one can use
the fact that $
\int_{-\infty }^{\infty }dX_{i}\int_{-\infty }^{\infty
}dp_{i}f(X_{i},p_{i},t_{i})=1$, which is as a consequence of ${\rm Tr}\rho
_{r}(t_{i})=1$ to obtain,
\begin{equation}
\bar{K}\int {\cal D}\xi \;P_{Q}[\xi ;t)=\bar{K}\int {\cal D}\xi \;P[\xi ]=1,
\label{1.19}
\end{equation}
which determines $\bar{K}$.

Several remarks are in order here. The functional $P_{Q}[\xi ,X[\xi
;X_{i},p_{i}) ;t)$ is
always real, but in general it will not be positive definite and, thus, will
not really correspond to the probability density functional for a classical
stochastic process. This is the meaning that must be associated to the
stochastic process in the Langevin-like equation (\ref{langevin}). This
situation is, in fact, analogous to that for the Wigner function but applied
here to distribution functionals. Note that this is in contrast to the
linear case studied in Ref. \cite{CRV00}, where the source of the Langevin
equation really corresponded to a stochastic process (with a positive
probability density functional).

We emphasize again that if we have a cubic potential for the system,
$V(x)=-(\lambda /6)x^{3}$, and keep up to quadratic order in $g$, we have
explicitly
\begin{equation}
F[\Delta ,X]=e^{-\frac{1}{2\hbar }\Delta \cdot N\cdot \Delta }
e^{-\frac{i} {\hbar }\int dt{\frac{\lambda}{24}}\Delta ^{3}}
\text{,}  \label{cf1}
\end{equation}
where the noise kernel $N$ and the kernel $H$, which appeared in $L$ above,
are both quadratic in $g$; note that there is no dependence on $X$ in this
case, and thus $P_Q[\xi ;t)$ defined in (\ref{1.18a}) does not depend
on the initial conditions $(X_i,p_i)$.
Hu, Paz and Zhang \cite{hu93} obtained the master
equation for the particular case in which the nonlinear potential of the
system is also treated perturbatively in $\lambda $, which was considered to
be of the same order as $g$. Here, however, our
result is exact in $\lambda $,
and to leading order in $\hbar $. This fact will turn out to be important
since in Sec. \ref{iii} it will be crucial to consider solutions of the
classical equations of motion which are nonperturbative in $\lambda $ thus
reflecting their strong nonharmonicity.

\subsection{The master equation}

The expression (\ref{wigner}) of the reduced Wigner function and the
Langevin-like equation (\ref{langevin}) can be used to derive the master
equation for $f$ as a formal Fokker-Planck equation. The derivation is
usually handled using Novikov's formula when the stochastic process is
Gaussian \cite{novikov65,sancho80,CRV00}. Here, however, this is not the
case for $P_Q$ and we have to work from the beginning.

To obtain the equation of motion for the Wigner function, we derive both
terms of eq. (\ref{wigner}) with respect to time. Observe that $P_{Q}$
depends explicitly on time, therefore
\begin{equation}
\frac{\partial }{\partial t_{f}}f(X_{f},p_{f},t_{f})=A+B,
\end{equation}
where
\begin{equation}
A=\bar{K}\int dX_{i}dp_{i}\int {\cal D}\xi \;P_{Q}[ \xi;
t_{f}^{-}) \;\left( \frac{\partial }{\partial t_{f}}\left[ \delta
\left( p_{f}-\dot{X}\left( t_{f}\right) \right) \delta \left( X_{f}-X\left(
t_{f}\right) \right) \right] \right) f(X_{i},p_{i},t_{i}),
\end{equation}
(we write $t_{f}^{-}$ in $P_{Q}$ to emphasize that the dependence on $t$ is
taken care of explicitly by the $B$ term) and
\begin{equation}
B=\bar{K}\int dX_{i}dp_{i}\int {\cal D}\xi \;\left( \frac{\partial }{
\partial t_{f}}P_{Q}[ \xi ;t_{f}) \right) \;\delta \left( p_{f}-
\dot{X}\left( t_{f}\right) \right) \delta \left( X_{f}-X\left( t_{f}\right)
\right) f(X_{i},p_{i},t_{i}).
\end{equation}

Let us analyze the $B$ term first. Since
\begin{equation}
\frac{\partial }{\partial t_{f}}P_{Q}[\xi ;t_{f})=\frac{i}{\hbar }
V^{(nl)}[-i\hbar \frac{\delta }{\delta \xi \left( t_{f}\right) }
,X_{f}]P_{Q}[\xi ;t_{f}^{-}),
\end{equation}
integrating by parts we find,
\begin{equation}
B=\bar{K}\int dX_{i}dp_{i}\int {\cal D}\xi \;P_{Q}[ \xi ;t_{f})
\left[ \frac{i}{\hbar }V^{(nl)}[i\hbar \frac{\delta }{\delta \xi \left(
t_{f}\right) },X_{f}]\left( \delta \left( p_{f}-\dot{X}\left( t_{f}\right)
\right) \delta \left( X_{f}-X\left( t_{f}\right) \right) \right) \right]
f(X_{i},p_{i},t_{i})
\end{equation}
We are only interested in derivatives taken at $t_{f}$, when
\begin{equation}
\frac{\delta X(t_{f})}{\delta \xi \left( t_{f}\right) }=0,
\qquad \frac{
\delta \dot{X}(t_{f})}{\delta \xi \left( t_{f}\right) }=1,
\end{equation}
without further dependence on  $\xi \left( t_{f}\right)$. This can be seen
from the fact that $\delta X(t)/\delta \xi (t^{\prime })$ satisfies
$((\bar{L
}+H)\cdot \delta X/\delta \xi )(t,t^{\prime })=\delta (t-t^{\prime })$
with the $X(t)$ which  appears in $\bar{L}$ fixed,
$\bar{L}$ is the integro-differential operator $\bar{L}
(t,t^{\prime })\equiv \left( d^{2}/dt^{2}+\left. \partial ^{2}V/\partial
X^{2}\right| _{X^{(0)}}\right) \delta (t-t^{\prime })$;
see Eq. (\ref{langevin}).
The solution is $\delta
X(t)/\delta \xi (t^{\prime })=G_{ret}(t,t^{\prime })$, which is the retarded
propagator corresponding to the linear operator $(\bar{L}+H)$ with $X(t)$
fixed. So the final result is
\begin{equation}
B=\frac{i}{\hbar }V^{(nl)}[-i\hbar \frac{\delta }{\delta p_{f}}
,X_{f}]f(X_{f},p_{f},t_{f}).
\end{equation}

Concerning $A$, we find
\begin{equation}
A=-\bar{K}\int dX_{i}dp_{i}\int {\cal D}\xi \;P_{Q}[ \xi
;t_{f}^{-}) \;\left[ \left( \frac{d X(t_{f})}
{d t_{f}}\frac{
\partial }{\partial X_{f}}+\frac{d \dot{X}(t_{f})}
{d t_{f}}\frac{
\partial }{\partial p_{f}}\right) \left[ \delta \left( p_{f}-\dot{X}\left(
t_{f}\right) \right) \delta \left( X_{f}-X\left( t_{f}\right) \right)
\right] \right] f_{i},
\end{equation}
and  reading  the derivatives from Eq. (\ref{langevin}) we can write
$A=A_{1}+A_{2}+A_{3}.$ The first term is simply
\begin{equation}
A_{1}=\{H_{s},f\},
\end{equation}
where $\{H_{s},f\}=-p(\partial f/\partial X)+V^{\prime }
(\partial f/\partial
p)$ is the Poisson bracket with $H_{s}=p^{2}/2+V(X)$
the system Hamiltonian. The second term is
\begin{equation}
A_{2}=-\frac{\partial }{\partial p_{f}}\bar{K}\int dX_{i}dp_{i}\int {\cal D}
\xi \;P_{Q}[ \xi ;t_{f}^{-}) \left[
\int_{t_i}^{t_f} dt^{\prime
}H(t_f,t^{\prime })X(t^{\prime })\right] \left[ \delta \left( p_{f}-\dot{X}
\left( t_{f}\right) \right) \delta \left( X_{f}-X\left( t_{f}\right) \right)
\right] f_{i},
\end{equation}
to lowest order in $\hbar ,$ we are entitled to replace $X(t^{\prime })$
inside the non-local term by a solution of the {\it classical} equations
of motion with the given Cauchy data $X_{f}$ and $p_{f}$. We shall call this
procedure of substitution of the classical trajectories into the terms which
are already of order $\hbar $ ``reduction of order''. We may then extract
the non-local term from the integral to get the simpler form
\begin{equation}
A_{2}={\frac{\partial }{\partial p}}[\Gamma (X,p,t)f],
\end{equation}
where $\Gamma (X,p,t)=-\int_{t_i}^{t}dt^{\prime }
H(t,t^{\prime })X(t^{\prime })$.
Finally, the third term is
\begin{equation}
A_{3}=-\frac{\partial }{\partial p_{f}}
\bar{K}\int dX_{i}dp_{i}\int {\cal D}
\xi \;P_{Q}[ \xi :t_{f}^{-}) \;\xi \left( t_{f}\right) \delta
\left( p_{f}-\dot{X}\left( t_{f}\right) \right) \delta \left( X_{f}-X\left(
t_{f}\right) \right) f_{i}.
\label{1.A3}
\end{equation}
To compute this term, we note from (\ref{1.17a}) and (\ref{1.18a}) that
\begin{equation}
\xi \left( t_{f}\right) P_{Q}[ \xi ;t_{f}^{-}) =\left\{ \exp
\left[ \frac{i}{\hbar }\int^{t_{f}^{-}}dt^{\prime }\;
V^{(nl)}[-i\hbar \frac{
\delta }{\delta \xi },X]\right] \right\} \xi \left( t_{f}\right) P\left[ \xi
\right],
\end{equation}
and since $P[\xi]$ is Gaussian, we may use the identity
(Novikov's formula):
\begin{equation}
\xi \left( t_{f}\right) P_{Q}[ \xi :t_{f}^{-}) =-\hbar
\int_{t_i}^{t}dt^{\prime }N(t,t^{\prime })
\frac{\delta }{\delta \xi (t^{\prime
})}P_{Q}[ \xi ;t_{f}^{-}).
\end{equation}
Integrating Eq. (\ref{1.A3}) by parts and
after further simplification, where
we explicitly assume that, as in the case of the cubic potential,
$P_Q$ is independent of $X$ (in this way we can commute the exponential
of $V^{(nl)}$ in $P_Q$ with the functional derivative with
respect to $\xi$), we obtain
\begin{equation}
A_{3}=\frac{\partial }{\partial p_{f}}\left\{ N,f\right\},
\end{equation}
where $N\equiv \hbar \int dt^{\prime }\,
N\left(t,t^{\prime }\right) X\left(
t^{\prime }\right) $, and we have applied once again a reduction of order
prescription. The details of this calculation are given in Appendix A.

To summarize, and using the explicit form of $V^{\left( nl\right) }$
for the cubic potential, we
obtain the following dynamical equation for
the reduced Wigner function (master
equation):
\begin{equation}
\frac{\partial f}{\partial t}=\left\{ H_{s},f\right\} +{\frac{\partial }
{\partial p}}[\Gamma (X,p,t)f]+\frac{\partial }{\partial p_{f}}\left\{
N,f\right\} -\hbar ^{2}\frac{\lambda }{24}\frac{\partial ^{3}f}{\partial
p^{3}}.  \label{1.22}
\end{equation}
If the system were isolated, the master Eq. (\ref{1.22}) would reduce to
\begin{equation}
\frac{\partial W}{\partial t}=\left\{ H_s,W\right\}
 -\hbar ^{2}\frac{\lambda
}{24}\frac{\partial ^{3}W}{\partial p^{3}}\text{,}  \label{master}
\end{equation}
where $W$ is the Wigner function of the closed system, $H_s$ its Hamiltonian
and the curly brackets are the Poisson brackets. This equation is exactly
equivalent to von Neumann's equation for
the density matrix of a one-dimensional quantum
mechanical system with a cubic potential $V(x)=-(\lambda/6)x^{3}$. Note that
the term with the third derivative with respect to the momentum is the
responsible for tunneling when properly combined with the otherwise
classical dynamics generated by the term corresponding to the Poisson
bracket, {\it i.e.}, if this term were not present, the evolution of the
Wigner function would be entirely equivalent to that of a classical ensemble
in phase space.

There is a theorem by Pawula \cite{risken89} which states that a
diffusion-like equation such as Eq. (\ref{master}) should have up to second
order derivatives at most, or else an infinite Kramers-Moyal expansion, for
non-negative solutions $W(x,p,t)$ to exist. The equation for the Wigner
function circumvents the implications of the theorem since it need not be
everywhere-positive. Even if we have an everywhere-positive Gaussian Wigner
function at the initial time, the evolution generated by an equation such as
Eq. (\ref{master}) will not keep it everywhere-positive. This can be
connected with the fact that the Wigner function could be interpreted as the
distribution function associated to an ensemble of solutions satisfying the
Langevin-like equation (\ref{langevin}) with a generalized stochastic source
$\xi (t)$ with a non-positive probability density functional $P_{Q}[\xi ]$.
Thus, here we see the essential role played by the non-positivity of the
Wigner function in a genuinely quantum aspect such as tunneling. In other
aspects such as in quantum coherence this role is well known \cite{zurek}.

It is important to stress the following points. First, we are not assuming
that the stochastic trajectories described by the Langevin equation are real
trajectories; although they may be if there is decoherence \cite{Hartle}.
Second, although we use the initial Wigner function to weight the initial
conditions, we do not asume it is a probability distribution function. Our
interest in the Wigner function is that we shall use the fact that it
acquires a substantial nonzero average beyond the basin of attraction of the
false vacuum as a signal that tunneling has occurred. This application is
valid even if the Wigner function itself cannot be understood as a
probability distribution function, because the distributions computed from
the Wigner functions, such as $\int_{-\infty }^{\infty }dX\;f(X,p)$ or $
\int_{-\infty }^{\infty }dp\;f(X,p)$ are true probability distributions, and
$\int_{-\infty }^{\infty }dX\;(p/M)f(X,p)$ is a true probability flux \cite
{wigner}.

\subsection{Dynamics of the distribution function $f$}

To compute the tunneling probability from equation (\ref{master}) is
possible using, for instance, a WKB approximation scheme
\cite{Banks,Gleiser}.  Alternatively one may use the
instanton method \cite{Coleman} which gives
a simple answer in this case. Our main interest in this paper is not this
contributon to tunneling but rather to compute the effect due to the back
reaction of the environment. To simplify our derivation we shall assume that
the time scales are different, that the dominant term is the back-reaction
effect and, thus, we will neglect the third derivative term in the master
equation (\ref{1.22}). More precisely, we will use the evolution equation
for the distribution function $f$
\begin{equation}
\frac{\partial f}{\partial t}=\left\{ H_{s},f\right\} +{\frac{\partial }
{\partial p}}[\Gamma (X,p,t)f]+\frac{\partial }{\partial p_{f}}\left\{
N,f\right\}   \label{lange3}
\end{equation}
This equation describes an ensemble of points evolving according to the
dynamics
\begin{equation}
\dot{X}=p,\qquad \dot{p}=-V^{\prime }\left( X\right) +\int_{-\infty
}^{t}dt^{\prime }\;H\left( t-t^{\prime }\right) X\left( t^{\prime }\right)
+\xi ,  \label{lange2}
\end{equation}
where $H$ is the dissipation kernel introduced in Eq. (\ref{1.12}), and $\xi
$ is a {\it Gaussian} noise described by the noise kernel $N$ introduced in
(\ref{1.9}), as
\begin{equation}
\left\langle \xi \left( t\right) \xi \left( t^{\prime }\right) \right\rangle
=\hbar N\left( t-t^{\prime }\right) .
\label{noise}
\end{equation}

Although we set up the initial conditions at $t_{i}=0$, we extend the lower
integration limit in Eq. (\ref{lange2}) to $-\infty $ for computational
purposes. The approximation is, nevertheless, justified since the
characteristic frequencies of the environment, $\omega _{j}\gtrsim 1$
(recall that the system particle has unit mass), are much larger than the
typical decay rate of the initial false vacuum state. In other words, the
characteristic time scale for the environment dynamics $t\lesssim 1$ and
hence, the typical ``correlation time'' for the dissipation and noise
kernels, is much smaller than the typical decay time. The contribution to
the integration interval $(-\infty ,0)$ is, therefore, relatively small.

According to Eqs. (\ref{1.10}) and (\ref{1.11}) these kernels admit the
following representation,
\begin{eqnarray}
N\left( t-t^{\prime }\right) &=&\frac 12\left\langle \left\{ \Xi \left(
t\right) ,\Xi \left( t^{\prime }\right) \right\} \right\rangle -\left\langle
\Xi \left( t\right) \right\rangle \left\langle \Xi \left( t^{\prime }\right)
\right\rangle,  \label{noiseker} \\
H\left( t-t^{\prime }\right) &=&\frac i\hbar \left\langle \left[ \Xi \left(
t\right) ,\Xi \left( t^{\prime }\right) \right] \right\rangle \theta \left(
t-t^{\prime }\right),  \label{dissiker}
\end{eqnarray}
where $\Xi\equiv\sum_j \Xi_j$ and where we must keep in mind the need for
regularization of the kernel $H$ as defined above. Let us take the Fourier
transforms,
\begin{eqnarray}
\left\langle \left[ \Xi \left( t\right) ,\Xi \left( t^{\prime }\right)
\right] \right\rangle &=&\int \frac{d\omega }{2\pi }\;e^{-i\omega \left(
t-t^{\prime }\right) }\hbar \omega \gamma \left( \omega \right),
\label{disfourier} \\
N\left( t-t^{\prime }\right) &=&\int \frac{d\omega }{2\pi }\;e^{-i\omega
\left( t-t^{\prime }\right) }\nu \left( \omega \right).  \label{noisefourier}
\end{eqnarray}
For an environment initially in thermal equilibrium at temperature $
T=\beta^{-1}$ the functions $\gamma $ and $\nu $ will be related through the
fluctuation-dissipation theorem
\begin{equation}
\nu \left( \omega \right) =\left[ \frac 12+f_0\left( \omega \right) \right]
\hbar \left| \omega \right| \gamma \left( \omega \right),\qquad f_0\left(
\omega \right) =\left( e^{\beta \hbar \left| \omega \right| }-1\right) ^{-1},
\label{fdt}
\end{equation}
which is a consequence of the KMS formula. Here we shall consider the zero
temperature case only.

In order to compute the memory terms in equation (\ref{lange2}), it is
convenient to parametrize the trajectories by their initial conditions at
time $t_i=0$. These trajectories may be written in terms of the action-angle
variables $J$ and $\theta $ associated to the classical potential, which we
assume it has a potential well bounded by a finite potential barrier. In
other words, we are using $J$ and $\theta $ as Lagrangian coordinates,
identifying a given trajectory, while $X$ and $p$ are like Eulerian
coordinates, identifying where the trajectory is at a given time. The action
variable is defined by $J=(1/2\pi)\oint pdX$; since $p$ can be written in
terms of $\dot X$ and the system Hamiltonian $H_s$, substitution into the
equation defining $J$ and inversion implies that $H_s=H_s(J)$, and $
dH_s/dJ=\Omega(J)$ is the frequency of the motion. The angle variable $
\theta $ changes from $0$ to $2\pi$ and satifies the
equation of motion $\dot\theta=\Omega$.

Since the kernel $H$ is already of order $\hbar$, in the memory term of (\ref
{lange2}) we must replace the trajectory $X\left( t^{\prime }\right)$ by a
solution of the classical equation of motion, in which case the
transformation to variables $\left( J,\theta \right) $ is canonical. For
fixed $J$, the classical trajectory is periodic, $X(t^\prime)=X(t^\prime
+2\pi/\Omega)$; note that the motion is periodic inside the potential well
but at higher energies, near the top of the potential barrier (the
separatrix) when $J\to J_s$ , the motion ceases to be periodic and
$\Omega\to 0$. Thus we may write
\begin{eqnarray}
X\left( t^{\prime }\right) &=&\sum_ne^{in\left[ \theta +\Omega \left(
J\right) t^{\prime }\right] }X_n\left( J\right)  \label{fourier} \\
p\left( t^{\prime }\right) &=&i\Omega \left( J\right) \sum_ne^{in\left[
\theta +\Omega \left( J\right) t^{\prime }\right] }nX_n\left( J\right)
\nonumber
\end{eqnarray}
where $X_{-n}=X_{n}^{*}$, since $X$ is real. We then write the memory
dependent term as:
\begin{equation}
\Gamma =-\int_{-\infty }^tdt^{\prime }\;H\left( t-t^{\prime }\right) X\left(
t^{\prime }\right) =-\sum_nX_n\left( J\right) e^{in\left[ \theta +\Omega
\left( J\right) t\right] }\int \frac{d\omega }{2\pi }\;\frac{\omega \gamma
\left( \omega \right) }{\omega +n\Omega \left( J\right) -i\varepsilon },
\end{equation}
where we have used (\ref{dissiker}), (\ref{disfourier}) and (\ref{fourier}),
and that $\int_{0}^\infty du\exp(isu)=i/(s+i\varepsilon)$. Therefore we have
\begin{equation}
\Gamma \left( X,p,t\right) =-\sum_nX_n\left( J\right) e^{in\left[ \theta
+\Omega \left( J\right) t\right] }\gamma _n\left( J\right) ,\qquad \gamma
_n\left( J\right) =\int \frac{d\omega }{2\pi }\;\frac{\omega \gamma \left(
\omega \right) }{\omega +n\Omega \left( J\right) -i\varepsilon }
\label{gamma}
\end{equation}
Observe that although the Langevin equation is now local in time, it is not
necessarily ohmic. A similar manipulation of the last term in the master
equation  (\ref{lange3}) gives (see Appendix A)
\begin{equation}
N\left( J,\theta \right) =\sum_nX_n\left( J\right) e^{in\left[ \theta
+\Omega \left( J\right) t\right] }N_n\left( J\right), \qquad N_n\left(
J\right) =\int \frac{d\omega }{2\pi }\;\frac{\left[ -i\nu \left( \omega
\right) \right] }{\omega +n\Omega \left( J\right) -i\varepsilon }.
\label{ene}
\end{equation}

\subsection{Weak dissipation limit: averaging over angles}

So far, we have kept $f$ arbitrary. To study tunneling, however, we may
impose the additional condition that $f=f(J),$ and obtain a simpler equation
by averaging the Fokker-Planck equation over the angle variable $\theta $.
This approximation has been discussed by Kramers \cite{Kramers,risken89} as
prevailing in the weak dissipation limit. Recall that then $\{H_s,f\}=0$,
and that, for any phase space function $\Psi(\theta,J)$,

\begin{eqnarray*}
\oint d\theta \;\left\{ X,\Psi \right\} &=&\oint dX\;\left\{ \left. \frac{
\partial \Psi }{\partial J}\right| _\theta -\left. \frac{\partial X}
{\partial J}\right| _\theta \frac{\dot \Psi }{\dot X }\right\} \\
&=&\frac d{dJ}\oint dX\;\Psi =\frac d{dJ}\left[ \frac 1\Omega \oint d\theta
\;p\Psi \;\right],
\end{eqnarray*}
where we have used that for the classical trajectory $d\theta=\dot\theta
dt=\Omega dt$, and that $dX=\partial_\theta X\vert_J d\theta+ \partial_J
X\vert_\theta dJ$ from where by imposing $dX=0$ we can deduce
$\partial_J\theta\vert_X$.

Finally the Fokker-Planck equation (\ref{lange3}) becomes
\begin{equation}
\frac{\partial f}{\partial t}=\frac d{dJ}\left\{ {\cal N}\frac{df}{\Omega dJ}
+{\cal D}f\right\}  \label{fokker}
\end{equation}
where we introduced ${\cal D}$ and ${\cal N}$ as follows
\begin{eqnarray}  \label{Dcurly}
{\cal D}&\equiv&\frac 1\Omega \oint d\theta \;p\Gamma =i\sum_n\left|
X_n\left( J\right) \right| ^2n\gamma _n\left( J\right), \\
{\cal N}&\equiv&\oint d\theta \;p\frac{\partial N}{\partial \theta }=\Omega
\sum_n\left| X_n\left( J\right) \right| ^2n^2N_n\left( J\right).
\label{Ncurly}
\end{eqnarray}
Now observe that from Eq. (\ref{gamma}) and using that $1/(s+i\varepsilon)=
{\rm PV}(1/s) +i\pi\delta(s)$ we can write
\[
\gamma _n\left( J\right) ={\rm PV}\int \frac{d\omega }{2\pi }\;\frac{\omega
\gamma \left( \omega \right) }{ \omega +n\Omega \left( J\right) }-\frac i2
n\Omega \left( J\right) \gamma \left[ n\Omega \left( J\right) \right],
\]
but since $1/(\omega -n\Omega \left( J\right) )-1/(\omega +n\Omega \left(
J\right) )=2n\Omega/(\omega ^2-n^2\Omega ^2\left( J\right) ) $ the first
term of $\gamma_n$ above integrates to zero, so that only the imaginary term
contributes, and we finally have
\begin{equation}
{\cal D}=\frac \Omega 2\sum_n\left| X_n\left( J\right) \right| ^2n^2\gamma
\left[ n\Omega \right].  \label{Dcurlybis}
\end{equation}
A similar computation using expression (\ref{ene}) for $N_n$ leads to the
final expression for ${\cal N}$:
\begin{equation}
{\cal N}=\frac \Omega 2\sum_n\left| X_n\left( J\right) \right| ^2n^2\nu
\left[ n\Omega \right].  \label{Ncurlybis}
\end{equation}

\subsection{A rough estimate of the decay rate}

Eqs. (\ref{lange3}) and (\ref{fokker}) are the basic equations for the rest
of our analysis. The rest of the paper is devoted to the explicit
computation of the ${\cal D}$ and ${\cal N}$ functions in a field
theoretical problem, and to solving the dynamical equations therefrom.
However, we may already at this point make an educated guess about the
relationship between the decay rate predicted by these equations, and the
usual quantum estimates.

The point is that, since these are after all equations similar to those
discussed by Kramers \cite{Kramers}, we may obtain a rough estimate of the
transition amplitude by just plugging in Kramers' result for the flux. This
is made from Eq. (\ref{fokker}) which we may write
as a continuity equation $
\partial_t f +\partial_J K=0$ where the probability flux $K$ may be directly
identified from the equation. Then one looks for stationary solutions with
positive flux $K_0$, which must satisfy $({\cal N}/\Omega)\partial_J f+
{\cal D} f=-K_0$. From this equation one may
estimate (imposing that the particle
is in the potential well $\int^{J_s}f(J)dJ\leq 1$, where $J_s$ is $J$ at the
separatrix, {\it i.e.} the top of the potential barrier) using that
$dE=\Omega dJ$ the following upper bound for $K_0$
\[
K_0\sim e^{-B\left( E_s\right) },\qquad B=\int dE\;\left({\cal D}/{\cal N}
\right).
\]
At high temperature, we have $\nu =\gamma kT$, ${\cal N}=kT{\cal D}$,
$B=E/kT $, and
\[
K_0\sim {\rm exp}\left( -E_s/kT\right),
\]
where $E_s$ is the energy at the separatrix, as expected \cite{risken89}. At
low temperature, see Eq. (\ref{fdt}), $\nu =\hbar \left| \omega \right|
\gamma /2. $ Now, because the sums which define both ${\cal N}$ and ${\cal D}
$ are dominated by frequencies of the order of the curvature of the
potential around the unstable fixed point $\omega \sim \sqrt{\left|
V_s^{\prime \prime }\right| }$ (this being the timescale for the exponential
approach to the unstable fixed point), then ${\cal N}\sim \hbar \sqrt{\left|
V_s^{\prime \prime }\right| }{\cal D}/2,$ and thus
\begin{equation}
K_0\sim {\rm exp}\left( -2E_s/\hbar \sqrt{\left|
V_s^{\prime \prime }\right|}
\right).  \label{estimate}
\end{equation}

We may compare this estimate with the usual WKB tunneling amplitude given by
\begin{equation}
K_0(tunnel)\sim {\rm exp}\left( -\hbar ^{-1}\int dx\;\sqrt{2V\left( x\right)}
\right).  \label{tunnelestim}
\end{equation}
If the integral is dominated by the peak at the unstable fixed point $X_s$,
then $V\sim E_s-V_s^{\prime \prime }\left( X-X_s\right) ^2/2$ and the
Euclidean trajectory may be parametrized as $X\sim X_s-\sqrt{
2E_s/V_s^{\prime \prime }}\cos \theta $, $0\leq \theta \leq \pi $, which
gives $\int dx\;\sqrt{2V\left( x\right) }\sim \pi E_s/\sqrt{V_s^{\prime
\prime }}. $

We arrive at the remarkable conclusion that tunneling is comparable to the
noise effect from the environment. Of course, our coarse estimates are not
reliable for an accurate comparison and we must proceed to a more
quantitative account.

\section{Quantum fields as open systems}

\label{iii}

\subsection{The model and system-environment split}

We wish to study vacuum decay at zero and finite temperatures for a 3+1
quantum massive scalar field $\Phi$ with action
\begin{equation}
S\left[ \Phi \right] =\int d^4x\;
\left(-{\frac{1}{2}} \eta^{\mu\nu}\partial_
\mu\Phi\partial_\nu\Phi -{\frac{1}{2}} M^2\Phi ^2 + {\frac{1}{6}}
g\Phi^3\right),  \label{action}
\end{equation}
where the metric convention is $\eta^{\mu\nu}={\rm diag}(-1,1,1,1)$ ($
\mu,\nu=0,1,2,3)$. Although we keep $\hbar $ explicit, we set $c=1$. The
mass $M$ has units of $(length)^{-1}$, $\Phi $ has units of
$M\sqrt{\hbar }$ and $g$ of $M/\sqrt{\hbar }$.

As discussed in the introduction, we wish to focus on the behavior of the
long wavelength modes of the field. Let us split the field
$\Phi=\phi+\varphi $, where $\phi$ represents the long wavelength modes and
$\varphi$ the short wavelength modes. To do that
we introduce a length scale $\Lambda^{-1}$ which
will be suitably fixed for our problem. To define $\phi(x)$
we take a window function $W(x^\prime-x)$ centered at the point $x$
with a width $\Lambda^{-1}$ and covolute the field $\Phi$ with it
\begin{equation}
\phi(x)=\int d^3 x^\prime W(x^\prime-x)\Phi(x^\prime),  \label{4.2}
\end{equation}
then, of course, $\varphi(x)=\Phi(x)-\phi(x)$. In this way the field at each
point has two contributions one corresponding roughly to scales larger or of
order $\Lambda^{-1}$ and the other to smaller scales. This has its
correspondence in momentum space, we may define the Fourier transform of
$\Phi$ by
\begin{equation}
\Phi(x)=\int {\frac{d^3 k}{(2\pi)^{3}}} e^{i\vec k\cdot \vec x}
\Phi_{\vec k
}(t),  \label{4.3}
\end{equation}
the long wavelength modes now become $\phi_{\vec k}(t)=\tilde W(-\vec k
)\Phi_{\vec k}(t), $ where $\tilde W(\vec k)$ is the
Fourier transform of $W$,
and for the short wavelength modes we have $\varphi_{\vec k}= \Phi_{\vec k
}-\phi_{\vec k}$. It may be convenient to use a Gaussian window in this way
$\tilde W$ is also Gaussian with a with $\Lambda$, or some times it may be
more convenient to take a step function for the Fourier transform of the
window such as $\tilde W(\vec k)= \theta(\Lambda-k)$ where $k=|\vec k|$.

At this point $\phi$ is still a field, that is, it contains an infinite
number of degrees of freedom. It is convenient to reduce the system to a
single degree of freedom, such as we have discussed in section \ref{ii}. One
way to accomplish this is simply to enclose the field in a box of size $
\Lambda^{-1}$; see \cite{Lee}. The boundary conditons in this case would
introduce an undesired discrete spectrum for the modes, from a physical
point of view it is more satisfactory to proceed as follows. In a region of
volume $\Lambda^{-3}$ we define the average field
\begin{equation}
\bar\phi(t)=\Lambda^3\int_{\Lambda^{-3}} d^3 x\phi(x).  \label{4.4a}
\end{equation}
Note that if we introduce the function $\rho(k/\Lambda)=\Lambda^3
\int_{\Lambda^{-3}}\exp (-i\vec k\cdot\vec x)$, which satisfies $\rho(0)=1$,
then in momentum space we have $\bar\phi(t)=(2\pi)^{-3}\int d^3
k\rho(k/\Lambda)\phi_{\vec k}$ so that $\bar\phi$ is made up of the modes of
the field with $k\leq\Lambda$.

Now when the the fields $\phi$ and $\varphi$ are substituted into Eq. (\ref
{action}) the action will be decomposed in three parts. One involves the
field $\phi$ only, another the field $\varphi$ only, both with the same
functional dependence as the original action and a third interacting part
involves terms linear in $\varphi$ and the quadratic term $(1/2)g\phi
\varphi^2$ which comes from the cubic term in Eq. (\ref{action}). We will
approximate the different terms as follows: $\int_{\Lambda^{-3}} d^3 x
\phi^n(x)\simeq \Lambda^3\bar\phi^n$ for any integer $n$ and $
\int_{\Lambda^{-3}}d^3 x\varphi(x)\simeq 0$ since the short wavelength modes
should average to zero. Thus, generally we have $\int_{\Lambda^{-3}}d^3
x\phi^n f(\varphi)\simeq \bar\phi^n\int_{\Lambda^{-3}}d^3 x f(\varphi)$,
where $f$ is an arbitrary polynomic function of $\varphi$. Since we want to
focus in the single degree of freedom $\bar\phi(t)$, which is the averaged
field associated to a certain region of volume $\Lambda^{-3}$, it will be
convenient to restrict the volume integrals involving the field $\phi$ to
that volume.

Up to this point the scale $\Lambda$ is arbitrary, now we need to fix it.
Since the field $\phi$ is homogeneous in region considered the gradient
terms should be negligible in front of the mass term, $(\vec\nabla
\phi)^2\ll M^2\phi^2$ and this means that the momentum cutoff for the modes
of the field $\phi$ should be $k\ll M$, which implies that there is an upper
bound for the scale $\Lambda$: $\Lambda\leq M$. We also want to be able to
treat the field $\varphi$ perturbatively and that means that the field
should be stable, in this case the cubic term $\varphi^3$ will be a two-loop
order term and may be neglected to leading order in $\hbar$. We will see in
a moment that the condition for $\varphi$ to be stable implies a lower bound
for $\Lambda$: $\Lambda\geq M$. Thus for the system-environment split to be
consistent we need $\Lambda\sim M$ and the volume is $M^{-3}$.

Finally the action (\ref{action}) can be aproximated as a system-environment
interaction action $S[\Phi]\simeq S_s[\bar\phi]+ S_e[\varphi]+S_{int}[\bar
\phi,\varphi]$ where
\begin{eqnarray}  \label{4.5b}
S_s[\bar\phi]&=&M^{-3}\int dt\left( {\frac{1}{2}} \dot{\bar\phi}^2
-{\frac{1}{2}}M^2\bar\phi^2 +{\frac{1}{6}}g\bar\phi^3\right), \\
S_e[\varphi]&=& \int dt\int_{k\geq M} {\frac{d^3 k}{(2\pi)^3}}
\left[ {\frac{1}{2}}\dot \varphi_{\vec k}
\dot \varphi_{-\vec k}-{\frac{1}{2}}
\left(k^2+M^2\right) \varphi_{\vec k}\varphi_{-\vec k}\right], \\
S_{int}[\bar\phi,\varphi]&=& -{\frac{1}{2}}g\int dt \int_{M^{-3}}d^3x \bar
\phi \varphi^2,  \label{4.5c}
\end{eqnarray}
where we made a mode decomposition and integrated over the whole space
volume in the environment action. Now the potential of the system
\begin{equation}
V(\bar\phi)= {\frac{1}{2}}M^2\bar\phi^2-{\frac{1}{6}}g\bar\phi^3,
\label{4.6}
\end{equation}
has a stable fixed point at $\bar\phi =0$ and an unstable fixed point at $
\bar\phi =\phi _s\equiv 2g^{-1}M^2.$ The former corresponds to zero energy,
and the later to $E=E_s\equiv M^{-1}\phi _s^2/6$, in the volume $M^{-3}$.
For intermediate energies, we may have bound and unbound states. They are
separated by a potential barrier, which at zero energy extends from $\bar\phi
=0$ to $\bar\phi = \phi _{exit}\equiv 3\phi _s/2$ . At any given energy
there will be three classical turning points $\phi _L<0<\phi _R<\phi _s<\phi
_X$; as $E\rightarrow 0,$ $\phi _L$, $\phi _R\rightarrow 0$ and $\phi
_X\rightarrow \phi _{exit}$, while when $E\rightarrow E_s$, $\phi _R,\phi
_X\rightarrow \phi _s$ and $\phi _L\rightarrow -\phi _s/2.$ We are thus in
the situation described in Sec. \ref{ii}, consequently according to the
estimate at the end of last subsection, Eq. (\ref{estimate}), we expect here
that the tunneling rate will be $K_0\sim {\rm exp} \left( -\alpha M^2/\hbar
g^2\right)$ where $\alpha$ is a dimensionless parameter to be determined.
Since the system is the averaged field $\bar\phi$ the tunneling rate is now
per unit volume. When comparing with Sec. \ref{ii} note that the system
coordinate $x$ and frequency $\Omega_0$ in the QBM model correspond here to
the averaged field $\bar\phi$ and the field mass $M$, respectively; also
here there is only one coupling parameter $\lambda=g$.

For the environment modes to be stable as required we should have from the
previous actions $S_e$ and $S_{int}$ that $k^2+M^2-g\bar\phi\geq0$, and
since the maximum value that $\bar\phi$ may take at the barrier is $
\phi_s=2g^{-1}M^2$, it is clear that $k\geq M$ and this gives the lower
bound for $\Lambda$ mentioned above.

Let us now follow Sec. \ref{ii} and define the function $\Xi(t)$ which
appears in the interaction action which is a key function to construct the
dissipation and noise kernels, see $\Xi_j(t)$ in Eqs. (\ref{1.10}) and (\ref
{1.11}), or $\Xi(t)$ in Eqs. (\ref{noiseker}) and (\ref{dissiker}). Thus we
likewise define
\begin{equation}
\Xi(t)\equiv M^3\int_{M^{-3}}d^3 x \frac 12g\varphi ^2\left( x\right),
\label{4.7}
\end{equation}
and the interaction action becomes
$S_{int}=M^{-3} \int dt \bar\phi(t)\Xi(t)$
; note that $\Xi(t)$ has units of $M^3\sqrt{\hbar }$. By analogy with (\ref
{1.8}) the influence functional becomes
\begin{equation}
e^{iS_{IF}/\hbar }=\int D\varphi D\varphi ^{\prime }\rho _{ei}
\exp{\frac{i}{\hbar}}\left[ S_e[\varphi]
-S_e[\varphi ^{\prime}] -M^{-3} \int dt
\left(\bar\phi(t)\Xi(t)- \bar\phi^\prime(t)\Xi^\prime(t)\right)\right],
\label{4.8}
\end{equation}
where $\rho_{ei}$ is the environment density matrix at the initial time. As
in the previous section, see Eq. (\ref{1.12}), we keep only quadratic terms
and we can write
\begin{equation}
S_{IF}={\frac{1}{2}}M^{-3} (\bar\phi-\bar\phi^\prime)\cdot H\cdot (\bar\phi+
\bar\phi^\prime)+ {\frac{i}{2}}M^{-3} (\bar\phi-\bar\phi^\prime) \cdot
N\cdot (\bar\phi-\bar\phi^\prime).  \label{4.9}
\end{equation}

Thus we are now in the situation described in general
terms in Sec. \ref{ii},
and must complete the following steps: (a) identify the kernels $\gamma $
and $\nu $, (b) evaluate the Fourier transforms of $x$ and $p$ along a
classical trajectory, (c) compute the functions ${\cal D}$ and ${\cal N}$,
(d) solve the Fokker-Planck equation for $f$, or at least justify the
approximations already discussed, and (e) evaluate the decay rate. We shall
carry tasks (a), (b) and (c) in this section, and leave (d) and (e) for the
next.

We should emphasize that as remarked in the previous section, the kernel $H$
needs to be regularized and that this involves a mass renormalization, we
assume from now on that the mass $M$ which appears in the action as well as
the coupling parameter $g$ are the renormalized values. Also we emphasize
that we have chosen to reduce the low frequency part of the field to a
single degree of freedom only for simplicity. An inhomogeneous field could
be handled, for example, with the techniques
presented in Ref. \cite{mazenko}.

\subsection{The dissipation and noise kernels: $\gamma $ and $\nu $}

In order to find the noise and dissipation kernels, we use the
representations in Eqs. (\ref{noiseker}) and (\ref{dissiker}). To compute
the averages, observe that the environmental variables can be treated as a
free field. The details for the evaluation of the correlation $\langle \Xi
(t)\Xi (t^{\prime })\rangle $ are reproduced in Appendix B, see Eq. (\ref
{C.1}), where we find
\begin{equation}
\left\langle \Xi \left( t\right) \Xi \left( t^{\prime }\right) \right\rangle
=\left\langle \Xi \left( t\right) \right\rangle \left\langle \Xi \left(
t^{\prime }\right) \right\rangle +\frac{\hbar ^2g^2M^3}8\int_{p\geq M}\frac{
d^3p}{\left( 2\pi \right) ^3}\frac{e^{-2i\omega _p\left( t-t^{\prime
}\right) }}{\omega _p^2}\text{.}  \label{qcorr}
\end{equation}

One may compute the anticommutator and commutator operator required for the
kernels (\ref{noiseker}) and (\ref{dissiker} from this expression. Then
comparing with Eqs. (\ref{disfourier}) and (\ref{noisefourier}), and
recalling that the dissipation kernel must be corrected by a factor of $
M^{-3}$, in agreement with Eq. (\ref{4.9}), we obtain
\begin{eqnarray*}
\hbar \omega \gamma \left( \omega \right) &=&{\frac{\pi\hbar ^2g^2}{4}}
\int_{p\geq M}\frac{d^3p}{\left( 2\pi \right) ^3\omega _p^2}\left[ \delta
\left( 2\omega _p-\omega \right) -\delta \left( 2\omega _p+\omega \right)
\right] \\
&=&2{\frac{\pi \hbar ^2g^2}{4}}{\rm sign}\left( \omega \right) \int_M^\infty
\frac{4\pi p^2dp}{\left( 2\pi \right) ^3\omega _p^2}\delta \left( 2\omega
_p-\left| \omega \right| \right) \\
&=&\frac{\hbar ^2g^2}{16\pi }{\rm sign}\left( \omega \right) \sqrt{1-\frac{
4M^2}{\omega ^2}}\theta \left( \omega ^2-8M^2\right),
\end{eqnarray*}
which leads to
\begin{equation}
\gamma \left( \omega \right) =\frac{\hbar g^2}{16\pi \left| \omega \right| }
\sqrt{1-\frac{4M^2}{\omega ^2}}\theta \left( \omega ^2-8M^2\right).
\label{4.10}
\end{equation}

With a similar computation, for the noise kernel, we get
\begin{equation}
\nu \left( \omega \right) =\frac{\hbar ^2g^2M^3}{32\pi }\sqrt{1-\frac{4M^2}{
\omega ^2}}\theta \left( \omega ^2-8M^2\right).  \label{4.11}
\end{equation}
These kernels are related as required by the zero temperature
fluctuation-dissipation theorem, see Eqs. (\ref{fdt}). We observe that these
are the same as the kernels found in Ref. \cite{CH94} in a different
context. The analysis there substantiates our claim that particle creation,
and backreaction thereof, are the main mechanism for dissipation and noise
in this model.

\subsection{Classical trajectories}

Let us recall the basic definitions. The form of the classical potential
energy density Eq. (\ref{4.6}) suggests writing $\bar\phi =\phi _s\bar x
=2g^{-1}M^2\bar x$ so that $V(\bar\phi)=4g^{-2}M^6(\bar x^2/2-\bar x^3/3)$.
Let us introduce also a dimensionless time $\tau=tM $ and the classical
action density for the system, which is defined as $\bar S_s=M^3 S_s$, reads
\begin{equation}
\bar S_s=\frac{4M^5}{g^2}\int d\tau \;\left[ \frac 12\left( \frac{d\bar x}
{d\tau } \right) ^2-v(\bar x) \right],  \label{4D.1}
\end{equation}
where $v(\bar x)=\bar x^2/2-\bar x^3/3$. This action density includes
dimensionless variables only, the critical point is now $\bar x=1$,
corresponding to dimensionless energy $\varepsilon _s=1/6 $. Physical energy
densities are of course $E=\left( 4M^6/g^2\right) \varepsilon \equiv
6E_s\varepsilon .$


\begin{figure}[tbp]
\centering \leavevmode \epsfysize=6cm \epsfbox{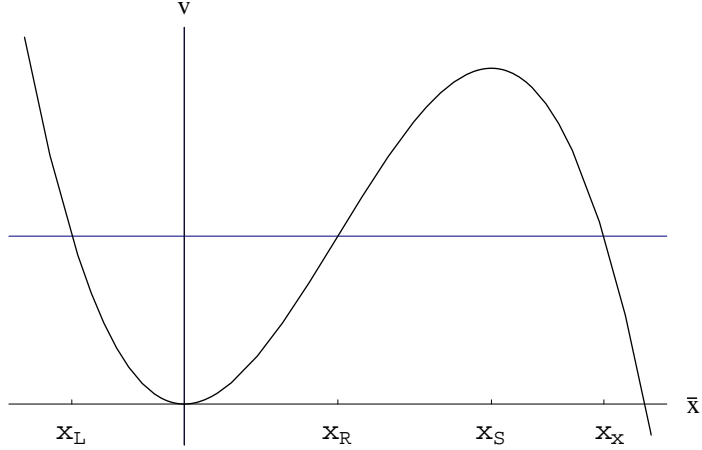}
\vspace{1cm} \caption[fig1]{Plot of the dimensionless potential $v(\bar{x})$,
as a function of the dimensionless variable $\bar{x}=\phi /\phi _s$. Note that
it exhibits a metastable minimum at $\bar{x}=0$ and an unstable maximum at
$\bar{x}=x_s$, where the value of the potential coincides with the energy
$\varepsilon _s$ of the separatrix trajectories.
The three turning points $x_L$, $x_R$ and $x_{\times }$ corresponding to any
energy $\varepsilon <\varepsilon _s$ have also been represented.}
\label{fig1}
\end{figure}


For any energy $\varepsilon <\varepsilon _s,$ we shall have three turning
points $x_L<0<x_R<1<x_{\times }$, see Fig. \ref{fig1}. The classical
orbits are expressed in terms of Jacobi elliptic functions with modulus
$\kappa $ given by
\begin{equation}
\kappa ^2=\frac{x_R-x_L}{x_{\times }-x_L},  \label{4D.2}
\end{equation}
which satisfies that $\kappa \rightarrow 0$ when $\varepsilon \rightarrow 0$
and $\kappa \rightarrow 1$ when $\varepsilon \rightarrow \varepsilon _{s}$.
The actual expression, computed in Appendix C, is
\begin{equation}
\bar x=x_L+\left( x_R-x_L\right) {\rm sn}^2\lambda \tau,  \label{4D.3}
\end{equation}
where $\lambda = \sqrt{(x_{\times }-x_L)/6}$. The function ${\rm sn}$ has
period $4K\left[ \kappa \right] $, and ${\rm sn} ^2$ has half of that, so
the dimensionless frequency is $\Omega =\lambda \pi /K$.

The Fourier coefficients of $\bar{x}(\tau )$, for $n\neq 0$, are computed in
Appendix D. They are given by
\begin{equation}
\bar x_n=-\frac{3n\Omega ^2}{\sinh \left( n\Omega K^{\prime }/\lambda\right)
}.  \label{foucoef}
\end{equation}

\subsection{The dimensionless Fokker-Planck equation}

Let us first write the noise and dissipation kernels in terms of
dimensionless variables. For the system Hamiltonian density, {\it i.e.}
$\bar H_s=M^3 H_s$, besides $\bar x$ and $\tau$
($\bar\phi =2g^{-1}M^2\bar x$
, $t=\tau /M$) we write also $p=2g^{-1}M^3\bar p$,
$V=4g^{-2}M^6v( \bar x)$
($V^{\prime }(\bar \phi) =2g^{-1}M^4v^{\prime }(\bar x)$),
thereby $\bar H_s=\left( 4g^{-2}M^6\right) {\bf h}_s$.
Note that this transformation is not
canonical but preserves the equations of motion, since $pd\bar\phi -\bar H
_sdt=\left( 4g^{-2}M^5\right) \left(\bar pd\bar x- {\bf h}_sd\tau \right)$,
see Ref. \cite{arnold}. The action variable also becomes $J=\left(
4g^{-2}M^5\right) j$, while the angles are, of course, unchanged. Since the
transformation is not canonical, it does not preserve phase volume: $dpd\bar
\phi = 4g^{-2}M^5\; d\bar pd\bar x.$ Therefore, we ought to define a new
distribution function $f\left(\bar\phi ,p,t\right) =\left( g^2/4M^5\right)
\bar f\left( \bar x,\bar p,\tau \right) $. Then we find $\partial f/\partial
t\rightarrow \left( g^2/4M^4\right) \partial \bar f/\partial \tau $, and $
\left\{ \bar H_s,f\right\} \rightarrow \left( g^2/4M^4\right)
\left\{ {\bf h}_s,\bar f\right\} _{\bar x,\bar p}.$
Therefore, instead of Eq. (\ref{lange3}), we have
\begin{equation}
\frac{\partial \bar f}{\partial \tau }=\left\{ {\bf h}_s,\bar f\right\} _{
\bar x,\bar p}+\frac g{2M^4}\frac \partial {\partial \bar p} \left[ \Gamma (
\bar\phi ,p,t)\bar f+\frac{g^2}{4M^5}\left\{ N,\bar f\right\} _{\bar x,\bar p
}\right].  \label{lange40}
\end{equation}

Now let us recall the definitions of $\gamma_n$ and $N_n$ in Eqs. (\ref
{gamma}) and (\ref{ene}) and let us introduce the new functions $\omega =M
\bar{\omega},$ $\gamma \left( \omega \right) =\left( \hbar g^2/M\right) \bar{
\gamma}\left( \bar{\omega}\right) ,$ $\gamma _n\left( J\right) =\hbar g^2
\bar{\gamma}_n\left( j\right) $ and $\bar\phi _n\left( J\right) = 2g^{-1}
M^2\; \bar x_n( j) $ to get $\Gamma =2\hbar gM^2\bar{\Gamma} \left( \bar x,
\bar p,\tau \right) $. On the other hand, let us call $\nu \left( \omega
\right) =\hbar ^2g^2M^3\bar{\nu}\left( \bar{\omega}\right) $, $N_n=\hbar
^2g^2M^3\bar{N}_n$ and $N=2\hbar ^2gM^5\bar{N}$ and we finally obtain the
following dimensionless form of the Fokker-Planck equation
\begin{equation}
\frac{\partial \bar{f}}{\partial \tau }=\left\{ {\bf h}_s,\bar{f}\right\} _{
\bar x,\bar p}+\frac 4\beta \frac \partial {\partial \bar p} \left[ \bar{
\Gamma}(\bar\phi ,p,t) \bar{f}+\frac 1\beta \left\{ \bar{N},\bar{f} \right\}
_{\bar x,\bar p}\right] ,  \label{lange4}
\end{equation}
where $\beta=4M^2/(\hbar g^2)$. In dimensionless variables, the energy scale
for the false vacuum is $1/(2\beta)$.

We may now proceed to write down the angle averaged Fokker-Planck equation (
\ref{fokker}) in dimensionless variables. This takes the form
\begin{equation}
\frac{\partial \bar{f}}{\partial \tau}=\frac 4\beta \frac d{dj}\left\{ \frac{
{\bf \bar{N}}}{\beta\bar\Omega} \frac{d\bar{f}}{dj}+{\bf \bar{D}}\bar{f}
\right\},  \label{fokker2}
\end{equation}
where ${\bf \bar{D}}=(\bar\Omega/2)\sum_n\left| \bar x_n\left( j\right)
\right| ^2n^2 \bar{\gamma}\left[ n\bar\Omega \right]$ and ${\bf \bar{N}}=(
\bar\Omega/2) \sum_n\left|\bar x_n\left( j\right) \right| ^2n^2\bar{\nu}
\left[ n\bar\Omega \right], $ and we have also introduced the dimensionless
$\bar\Omega=\Omega/M$. Using the explicit form of the Fourier coefficients
Eq. (\ref{foucoef}) and the expressions (\ref{4.10}) and (\ref{4.11}) for
the dissipation and noise kernels, $\gamma$ and $\nu$, we have explicitly
\begin{eqnarray}  \label{4D.4a}
{\bf \bar{D}}&=&\frac{9\bar\Omega }{16\pi }\sum_{n>0} \frac{\left( n\bar
\Omega \right) ^3}{\sinh ^2\left(n\bar\Omega K^{\prime }/ \lambda \right) }
\sqrt{1-\frac 4 {n^2\bar\Omega ^2}}\theta \left( n^2\bar\Omega ^2-8\right),
\\
{\bf \bar{N}}&=&\frac{9\bar\Omega }{32\pi }\sum_{n>0} \frac{\left( n\bar
\Omega \right) ^4}{\sinh ^2\left(n\bar\Omega K^{\prime }/ \lambda \right) }
\sqrt{1-\frac 4 {n^2\bar\Omega ^2}}\theta \left( n^2\bar\Omega ^2-8\right)
\label{4D.4b}
\end{eqnarray}

There are two relevant limiting cases. The first limit corresponds to the
separatrix energy, that is, when $\varepsilon \rightarrow \varepsilon _s$,
$\bar\Omega \rightarrow 0$, $\lambda \rightarrow 1/2,$ $K^{\prime
}\rightarrow \pi /2.$ We may write $\xi =n\bar\Omega $ and $\sum_n\sim
\left( 1/\bar\Omega \right) \int d\xi$ in the previous equations. Then after
numerical integration we obtain
\begin{equation}
{\bf \bar{D}}\equiv {\bf \bar{D}} _s\sim \frac 9{16\pi }2.441\dots\times
10^{-7}, \quad {\bf \bar{N}}\equiv {\bf \bar{N}} _s\sim \frac 9{32\pi }
7.381\dots\times 10^{-7},  \label{4D.5}
\end{equation}
which remain finite. The second limit corresponds to the bottom of the
potential, that is, when $\varepsilon \rightarrow 0$, $\bar\Omega
\rightarrow 1,\lambda \rightarrow 1/2$. Then $K^{\prime
}\sim(1/2)\ln(16\kappa^{-2})= (1/2)\ln[24(x_R-x_L)^{-1}]\sim (1/4)\ln
(72/\varepsilon)$, see Ref. \cite{GR}, formula 8.113.3, and $\sinh
^{-2}\left(n\bar\Omega K^{\prime }/\lambda \right) \sim 4\exp(-4nK^{\prime
})\sim 4(\varepsilon/72) ^n $ and the expressions for ${\bf \bar{D}}$ and
${\bf \bar{N}}$ take the values
\begin{equation}
{\bf \bar{D}}\sim \frac 9{4\pi }\sum_{n=3}n^3\left( \frac \varepsilon {72}
\right) ^n\sqrt{1-\frac 4{n^2}}\sim b\varepsilon ^3,\quad {\bf \bar{N}}\sim
\frac 9{8\pi }\sum_{n=3}n^4\left( \frac \varepsilon {72} \right) ^n\sqrt{1-
\frac 4{n^2}}\sim a\varepsilon ^3,  \label{4D.6}
\end{equation}
where $a$ and $b$ are numerical coefficients that can be read from these
expressions. Note that the rapid decay of these functions as $\varepsilon
\rightarrow 0$ is due to the presence of a threshold, encoded in the theta
functions of Eqs. (\ref{4.10}) and (\ref{4.11}), which enforce that $n^2\bar
\Omega^2> 8$. At finite temperature, where the self-energy of the
fluctuations remains complex even on-shell (when higher loops are included,
of course), ${\bf \bar{D}}$ and ${\bf \bar{N}}$ fall like $\varepsilon$, as
in Kramers' original analysis \cite{Kramers}.

\section{Environment induced decay rates}

\label{iv}

We may now conclude our task of finding the decay rates as described by the
Fokker-Planck equation. In agreement with our weak dissipation assumptions
we shall assume that we may average over angles, and restrict our analysis
to the Fokker-Planck equation (\ref{fokker2}); a related analysis of
angle-dependent solutions is given in Appendix 6 of Ref. \cite{CV}. Since
the equation is linear in the Wigner function, it is convenient to first
seek the normal modes, namely, solutions with a simple (exponential)
dependence on time. The desired solution will then be reconstructed as a
superposition of normal modes \cite{risken89,Gardiner}.

Assuming then that $\bar{f}=e^{-r\tau}F\left( j\right) $, we get from (\ref
{fokker2}) the time independent equation
\begin{equation}
LF+rF=0,\qquad L=\frac 4\beta \frac d{dj}\left( \frac{{\bf \bar{N}}} {\beta
\bar\Omega} \frac d{dj}+{\bf \bar{D}}\right).  \label{5.1}
\end{equation}
This equation may be written as a conservation equation $LF=d\bar\Phi/dj$,
where $\bar\Phi$ is the flux defined by
\begin{equation}
\bar\Phi =\frac 4\beta \left( \frac{{\bf \bar{N}}} {\beta\bar\Omega}
\frac{dF}{dj}+{\bf \bar{D}}F\right).  \label{5.2}
\end{equation}
Observe that for any $\varepsilon >\varepsilon ^{\prime }>0$ we have the
following relationship between $F$, $\bar\Phi $ and $r$:
\begin{equation}
r=\frac{\bar\Phi \left( \varepsilon \right) -\bar\Phi \left( \varepsilon
^{\prime} \right) }{\int_{\varepsilon ^{\prime }}^\varepsilon Fdj}.
\label{firstintegral}
\end{equation}

Let us now introduce the unnormalized equilibrium solution $F_B=\exp (
-\beta \int dj\;\bar\Omega {\bf \bar{D}}/{\bf \bar{N}}), $ then Eq. (\ref
{5.1}) can be written as
\begin{equation}
\frac d{dj}\left[ \rho ^2\frac d{dj}\left( \frac F{F_B}\right) \right] +rF=0,
\label{SL}
\end{equation}
where $\rho ^2\left( j\right) = 4{\bf \bar{N}}F_B/(\bar\Omega \beta ^2)$.
Multiplying by $F/F_B$ and integrating we see that $r$ must be non-negative,
as expected. From the mathematical point of view, this is an eigenvalue
problem of the Sturm-Liouville type, and we may handle it in the usual way
\cite{Courant}. Let us begin by analyzing the solutions with $r\neq 0$,
which are the decaying solutions.

\subsection{Decaying solutions}

Let us seek a solution $F_K$ of Eq. (\ref{SL}) with nonzero $r$. Since later
on we shall be interested in the long time behavior of solutions, we may
focus on the range of small values of $r$. For concreteness, let us assume
$r\beta ^3/3a\ll 1$. Let us introduce $h$ by $F_K=(h/\rho) F_B=(\beta h/2)
\sqrt{\bar\Omega F_B/{\bf \bar{N}}} $ and write Eq. (\ref{SL}) and the flux
as
\[
\frac d{dj}\left[ \rho ^2\frac d{dj}\left( \frac h\rho \right) \right] +
\frac{r\bar\Omega \beta ^2}{4{\bf \bar{N}}}\rho h=0, \qquad \bar\Phi =-\rho
^2\frac d{ dj}\left( \frac h\rho \right) =-( \rho h^{\prime }-h\rho
^{\prime}).
\]
Expanding the second derivatives $d_j[ \rho ^2d_j( h/\rho)] = d_j( \rho
h^{\prime }-h\rho ^{\prime }) =\rho h^{\prime \prime }-h\rho ^{\prime \prime
}, $ where $d_j$ stands for the derivative with respect to $j$, we obtain
\begin{equation}
h^{\prime \prime }+\left(\frac{r\bar\Omega \beta ^2} {4{\bf \bar{N}}}
-\frac{\rho ^{\prime \prime }}\rho \right) h=0.  \label{5A.1}
\end{equation}
We have two regimes in this equation: for large $\varepsilon $ (but not
exponentially close to $\varepsilon _s=1/6$), $\rho $ is dominated by $F_B$,
and $\rho ^{\prime \prime }/\rho \sim \beta ^2\bar\Omega ^2{\bf \bar{D}}^2/(4
{\bf \bar{N}}^2)$ dominates the $r$ term, which is negligible. For $
\varepsilon \rightarrow 0$, on the other hand,
$\rho \sim \varepsilon ^{3/2}$,
$\rho ^{\prime \prime }/\rho \sim \left( 3/4\right) \varepsilon ^{-2}$,
and the $r$ term is the leading term. The transition occurs for $\varepsilon
=\varepsilon ^{*}\sim r\beta ^2/(3a),$ which by assumption is much smaller
than the energy scale $1/2\beta $ typical of the false vacuum.

For large $\varepsilon$, the equation $h^{\prime \prime }-(\rho ^{\prime
\prime }/\rho) h=0 $ admits of course the solution $h=\rho $, which gives
back the equilibrium solution, {\it i.e.} $F_K=F_B$. To find the second
solution we may write $h=K_r\rho \sigma $ (here $K_r$ is a constant) to get
$\sigma ^{\prime \prime }/\sigma ^{\prime }=-2\rho ^{\prime }/\rho$ which
implies $\sigma^\prime=-\rho^{-2}.$ This second solution corresponds to a
constant flux $\bar\Phi =K_r$, and also is the negative exponential WKB
solution. For small $\varepsilon$, the equation $h^{\prime \prime }+(r\beta
^2/4a\varepsilon ^3)h=0 $ has solutions
\begin{equation}
h=C_r\sqrt{\varepsilon }Z_1\left[ \frac{k_r}{\sqrt{\varepsilon }}\right] ,
\label{bessel}
\end{equation}
where $k_r^2=r\beta^2/a$, $Z$ is a Bessel function and $C_r$ is a constant
(observe that all solutions are bounded). As $\varepsilon \rightarrow 0,$
$F\sim \varepsilon ^{-3/4}\cos \left( k\varepsilon ^{-1/2}\right) $
but $\bar\Phi \sim \varepsilon ^{3/4}\sin
\left( k\varepsilon ^{-1/2}\right)
\rightarrow 0$.

Note that we obtain solutions for arbitrarily small values of $r.$ This
behavior, which is unlike that found by Kramers \cite{Kramers}, reflects the
existence of a threshold for dissipation: at arbitrarily low energies, the
motion is essentially harmonic, there is no dissipation, the fluctuations
switch off (in agreement with the fluctuation-dissipation theorem) and the
system requires an infinite time to climb out of the potential well.

\subsection{Hilbert space structure and normalization}

The structure of Eq. (\ref{SL}) suggests introducing an inner product
\begin{equation}
\left\langle f,g\right\rangle =\int_0^{j_s}\frac{dj}{F_B}\;f^{*}g,
\label{IP}
\end{equation}
where the star means complex conjugation. Then, if $F_r$ and $F_s$ are the
solutions corresponding to eigenvalues $r$ and $s$, we can write $(
r-s)\langle F_s,F_r\rangle= -\int_0^{j_s}dj\; F_B^{-1}[ F_s( LF_r) -( LF_s)
F_r]= \left. -F_B^{-1}[ F_s\bar\Phi _r-\bar\Phi _sF_r] \right| _0^{j_s}. $
Imposing the boundary conditions $F_r\left( j_s\right) =0,$ we get rid of
the contribution from the upper limit. In the lower limit, we may use the
asymptotic form of the Bessel functions to get $[ F_s\bar\Phi _r-\bar\Phi
_sF_r] ( \varepsilon ) \sim 2\pi^{-1}C_rC_s k_r^{-1/2}k_s^{-1/2}\sin [ (
k_r-k_s) \varepsilon^{-1/2}] $ which converges (weakly) to zero. It is
therefore natural to adopt the continuum normalization prescription
\begin{equation}
\left\langle F_s,F_r\right\rangle \equiv \delta \left( r-s\right).
\label{normalization}
\end{equation}

As in Landau and Lifschitz' analysis of the WKB wave functions
\cite{Landau}, this singular behavior is caused by the
oscillations as $\varepsilon
\rightarrow 0$. More precisely $\int_0 dj\;F_B^{-1}F_rF_s \sim
\pi^{-1}C_rC_sk_r^{-1/2}k_s^{-1/2} \int_0 d\varepsilon\; \varepsilon
^{-3/2}\cos [ ( k_r-k_s) \varepsilon ^{-1/2}] \sim 4C_r^2 k_r^{-1}\delta (
k_r-k_s) \sim 8aC_r^2 \beta ^{-2}\delta ( r-s), $ where the integral upper
limit is anything. We thus find the constants $C_r$
\begin{equation}
C_r\sim \beta /\sqrt{8a}.  \label{CR}
\end{equation}

To find the constants $K_r$ we should match our solutions accross the
transition point at $\varepsilon ^{*}.$ Without getting into details, see
Ref. \cite{Heading}, it is clear that in order of magnitude
\begin{equation}
K_r\sim \frac 1{\rho \left( \varepsilon ^{*}\right) \sigma \left(
\varepsilon ^{*}\right) }\sqrt{\frac 2{\pi k_r}}\frac{\beta \varepsilon
^{*3/4}}{\sqrt{8a}}\sim \frac{3^{3/4}}{4\sqrt{\pi }r^{1/4}\sigma \left(
\varepsilon ^{*}\right) \varepsilon ^{*3/4}}\text{.}  \label{KR}
\end{equation}

To find $\sigma \left( \varepsilon ^{*}\right) $, we recall that according
to our boundary conditions $\sigma \left( j_s\right) =0$. Therefore $\sigma
( \varepsilon ^{*}) =\int_{\varepsilon ^{*}}^{\varepsilon _s}dj \rho ^{-2}(
j). $ Let us split the range of integration in the two segments $\left[
\varepsilon ^{*},1/2\beta \right] $ and $\left[ 1/2\beta ,\varepsilon
\right] $. The contribution $I_{II}$ from the second segment is dominated by
the exponential growth of $F_B^{-1}$ near the separatrix, thus $I_{II}\sim
F_B( 1/2\beta)F_B^{-1}( \varepsilon _s) \sim \Delta ^{\prime -1}\exp ( \beta
{\bf \bar{D}} _s /6{\bf \bar{N}}_s) $ where $\Delta ^{\prime }\sim O( 1)$.
In the first segment, we find $\rho ^{-2}=\beta ^2/(4a\varepsilon ^3)$, so
it contributes as $I_I\sim (1/8)\beta ^2a^{-1}[ (\varepsilon
^{*})^{-2}-(1/4)\beta^{-2}] \sim (1/8)\beta ^2a^{-1}[ 9a^2\beta
^{-4}r^{-2}-(1/4) \beta ^{-2}] \sim (9/8)a\beta ^{-2}[ r^{-2}- (1/36)\beta
^2 a^{-2}]. $ Adding both contributions, and keeping only the leading terms,
we find
\[
\sigma \left( \varepsilon ^{*}\right) = \frac{9a}{8\beta ^2}\left( \frac 1{
r^2 }+\frac 1{\lambda ^2}\right),
\]
where $\lambda \sim \tilde{\Delta}\exp( -\beta {\bf \bar{D}}_s/12{\bf
\bar{N}}_s) $ with $\tilde{\Delta}\sim O\left( 1\right).$ So finally we have
\begin{equation}
K_r\sim K\frac{\lambda ^2r}{r^2+\lambda ^2},  \label{KR2}
\end{equation}
where $K$ is a $r$-independent constant.

\subsection{Decay of the false vacuum}

After all this work, the dynamical problem is now trivial. We are interested
in a time-dependent solution with an initial condition similar to the Wigner
function of the false vacuum, $F\equiv F_{FV}\sim 2\beta \exp \left( -2\beta
\varepsilon \right) $ \cite{Salman}. The solution is
\[
F\left( j,t\right) =\int_0^\infty dr\;e^{-rt}c_rF_r\left( j\right)
\]
where $c_r=\int_0^{j_s}dj\;F_B^{-1}F_rF_{FV}.$ Fortunately, we are
interested in the range of small $r$, where $F_r$ peaks at values much
smaller than $1/2\beta $. The only feature of $F_{FV}$ and $F_B$ that we
really need is that they are smooth there. Thus, $c_r\sim
F_{FV}(0)F_B^{-1}(0) \int_0^{j_s}dj\;F_r=2\beta K_r/r, $ (cfr. Eq. (\ref
{firstintegral})). Finally
\[
F\left( j,t\right) =2\beta K\lambda ^2\int_0^\infty dr\;\frac{e^{-rt}}{
r^2+\lambda ^2}F_r\left( j\right).
\]

To find the persistence probability, $P(t)$, we integrate $F(j,t)$ over the
potential well. Thus after using the previous result for $
\int_0^{j_s}dj\;F_r(j)$ and the value of $K_r$ from (\ref{KR2}) we obtain
\[
P\left( t\right) \sim 2\beta K^2\int_0^\infty dr\;\frac{\lambda ^4e^{-rt}}
{\left[ r^2+\lambda ^2\right] ^2}.
\]
The analytic expression for this integral is given in Ref. \cite{GR},
formula 3.355.1. When $\lambda t\sim 1$ it can be approximated with an
integral, the best fit is obtained when $P\left( t\right) \sim \exp
(-0.4\lambda t).$ For larger times, we have a crossover from exponential to
power law $1/t$ decay, as expected from quantum
mechanics (cfr. \cite{Mario}). To summarize, we have
proven that the Fokker-Planck equation leads to an
exponential decay of the false vacuum, with a decay rate $
t_d^{-1}\sim\lambda $:
\begin{equation}
t_d^{-1}\sim \Delta \exp\left({\ \frac{-{\cal D}_s} {12{\cal N}_s}}\beta
\right) ,  \label{final}
\end{equation}
with $\Delta\sim O(1)$. This equation may be compared to Eq.
(\ref{arrhenius}), here $\beta=4M^2/(\hbar g^2)$, as
defined in Eq. (\ref{lange4}), and the
ratio ${\cal D}_s/{\cal N}_s= \bar{{\bf D}}_s/\bar{{\bf N}}_s$ where
$\bar{{\bf D}}_s$ and $\bar{{\bf N}}_s$ are given in (\ref{4D.6});
see Eqs. (\ref
{Dcurlybis}), (\ref{Ncurlybis}) and (\ref{fokker2}) for the definitions of
the functions involved.

\subsection{Comparing with the instanton method}

Let us conclude by comparing the rate estimate of Eq. (\ref{final}) with the
estimate derived from the instanton method. The Euclidean action for our
model is
\begin{equation}
S_E\left[ \Phi \right] =\int d^4x \left( {\frac{1}{2}}\delta^{\mu\nu}
\partial_\mu\Phi\partial_\nu\Phi +{\frac{1}{2}}M^2\Phi^2-{\frac{1}{6}}
g\Phi^3 \right),  \label{eu-action}
\end{equation}
where $\delta^{\mu\nu}$ stands for the Euclidean metric. We are interested
in $SO\left( 4\right) $ symmetric instantons, which depend on all four
Euclidean coordinates but only through the Euclidean radius $
\rho=(x^2+y^2+z^2+t_E^2)^{1/2}$, where $t_E$ is the Euclidean time. Let us
scale $\Phi =\phi _sf\left( M\rho \right) $ to get the Euclidean action
(in $d$ dimensions) $S_d=\left( 4\pi /3\right) \hbar \beta I_d, $ where
\begin{equation}
I_d=\int d\rho\;\rho^{d-1}\left[ \frac 12 \left( f^{\prime }\right) ^2+\frac
12f^2- \frac 13f^3\right].  \label{id}
\end{equation}
Of course, computing the four-dimensional ($d=4$) instanton is not simple,
but we may approximate $I_4$ by $I_1$, in which case we may use the simpler
one-dimensional formula
\[
S=2\int_0^{3/2}dx\;\sqrt{2v(x) }=2\int dx\;\sqrt{x^2-\frac 23 x^3}=\frac 65
\]
All in all, the instanton prediction is $\ln t_d^{-1}\sim
-1.2\dots\times\beta$. On the other hand, the noise induced prediction, Eq.
(\ref{final}), is $\ln t_d^{-1}\sim \ln \lambda \sim ( -{\bf \bar{D}}_s/12
{\bf \bar{N}}_s) \beta \sim -0.05\dots\times\beta$. So in this simple case
the noise induced contribution dominates over the instanton contribution.

\section{Conclusions}

\label{vi}

In this paper we have studied the contribution to vacuum decay in field
theory as a consequence of the interaction between the long and
short-wavelength modes. We have seen that the dynamics of the
long-wavelength modes becomes diffussive in its interaction with the short
wavelength modes. On the one hand, there is dissipation of the
long-wavelength modes due to the excitation of the short-wavelength sector,
and in turn, that latter sector induces fluctuations into the first sector.
As a result there is a significant contribution to the total decay rate due
to activation, even at zero temperature.

A few remarks are in order. What we have shown is that if we consider the
field initially in a metaestble phase in a region of size $M^{-1}$ there is
a probabiliy per unit volume and unit time of decay to a stable phase given
by Eq. (\ref{final}). Of course, the size of the bubble formed is of order
$M^{-1}$, in our calculation this size is fixed. We cannot consider smaller
size regions because we could have not neglected the gradient terms of the
system in front of the mass terms, also we could not have considered larger
size regions because the environment modes would become unstable, their
evolution would become nonlinear and our perturbative treatment of the
influence functional would break down. However, the critical bubble size
that one obtains in first order phase transition in statistical physics or
in field theory using instanton methods is also of the order of $M^{-1}$.
This, in our opinion, makes this computation of interest, since one expects
that once the critical bubble is formed it will evolve in the usual way; see
for instance Ref. \cite{Coleman}. Had our calculation involved a size
smaller than the critical size then the bubbles formed could not grow and
would collapse, as then the energy of the bubble wall would overcome the
energy difference between the metastable phase and the bubble interior.

Another relevant point concerns the energy balance in the activation
mechanism of vacuum decay described here which involves noise and
dissipation. In Appendix E we show that the average power exchanged between
system and environment is zero: if some trajectories gain power through
noise, some others lose power through dissipation. Of course, the balance is
only statistical, but we must stress that this gain and loss process would
go on even if there were no separatrix and the system were in equilibrium.
The only reason why the system does not equilibrate in our problem is that
we remove those particles that reach the separatrix, as demanded by our
boundary condition there. We should say that the system equilibrates but we
assume that the true vacuum is very much deeper than the false vacuum, so
the equilibrium distribution vanishes inside the potential well. Note that
this is an expected result in vacuum bubble formation. Once the critical
bubble is formed the energy released in the conversion from false to true
vacuum is converted into energy of the growing bubble wall, so that the
energy balance is still zero, this last aspect however cannot be studied
with the present analysis.

\section{Acknowledgments}

\label{vii}

We are grateful to Daniel Arteaga, Leticia Cugliandolo, Jaume Garriga,
Rodolfo Id Betan, Bei-Lok Hu, Ted Jacobson and Jorge Kurchan for many
interesting discussions.

This work has been partially supported by Fundaci\'on Antorchas under grant
A-13622/1-21. E.C. acknowledges support from Universidad de Buenos Aires,
CONICET, Fundaci\'on Antorchas and the ANPCYT through Project PICT99
03-05229. A.R. and E.V. have also been supported by the CICYT Research
Project No. AEN98-0431, A.R. also acknowledges partial support by a grant
from the Generalitat de Catalunya, and E.V. also acknowledges support from
the Spanish Ministery of Education under the FPU grant PR2000-0181
and the University of Maryland for hospitality.

\section{Appendices}
\label{v}

\subsection{Appendix A}

Our problem is to compute an expression like
\begin{equation}
\int dt^{\prime }\,N\left( t-t^{\prime }\right) \left\langle \frac{\delta }
{\delta \xi \left( t^{\prime }\right) }R\left[ X\left( t\right) ,p\left(
t\right) \right] \right\rangle ,  \label{novikov}
\end{equation}
where $R$ is an arbitrary functional of $\xi (t)$ and $\left\langle
\dots \right\rangle $ denotes expectation value with respect to
$P_{Q}[\xi;t)$. Here, for simplicity, we will assume also
that $P_Q$ is independent
of $X$ as in the case of the cubic potential. In our
case, this leads to
\[
\tilde\Phi[\xi] =-\hbar\int dt^{\prime }\,N\left( t-t^{\prime }\right)
\left[ \frac \partial {\partial X}\left\langle \frac{\delta X\left( t\right)
}{\delta \xi \left( \tau \right) }\delta _X\delta _p\right\rangle +\frac
\partial {\partial p} \left\langle \frac{\delta p\left( t\right) }{\delta
\xi \left( \tau \right) } \delta _X\delta _p\right\rangle \right],
\]
where $\delta_X\equiv\delta(X(t)-X)$ and $\delta_p\equiv\delta(p(t)-p)$.

Within the ``reduction of order'' procedure, we substitute $\partial X\left(
\tau \right) /\partial X\left( t\right) $ and $\partial X\left( \tau \right)
/\partial p\left( t\right) $ for $\delta X\left( t\right) /\delta \xi \left(
\tau \right) $ and $\delta p\left( t\right) /\delta
\xi \left( \tau \right)$,
where the variations are understood in the sense of the result of coupling
a stochastic source to the classical equations of motion. The argument runs
as follows: we know that $\delta X\left( \tau ^{+}\right) /\delta \xi \left(
\tau \right) =0$ and $\delta p\left( \tau ^{+}\right) /\delta \xi \left(
\tau \right) =1$. On the other hand $\delta_{\xi(\tau)}X(\tau^+)=
\partial_{X(t)}X(\tau^+)\delta_{\xi(\tau)}X(t) +
\partial_{p(t)}X(\tau^+)\delta_{\xi(\tau)}p(t) $ and $\delta_{\xi(\tau)}p(
\tau^+)= \partial_{X(t)}p(\tau^+)\delta_{\xi(\tau)}X(t) +
\partial_{p(t)}p(\tau^+)\delta_{\xi(\tau)}p(t) $. By Liouville's theorem the
determinant of this two by two system is one, so indeed
\[
\frac{\delta X\left( t\right) }{\delta \xi \left( \tau \right) }=-\frac{
\partial X\left( \tau ^{+}\right) }{\partial p\left( t\right) },\qquad
\frac{\delta p\left( t\right) }{\delta \xi \left( \tau \right) }=
\frac{\partial X\left( \tau ^{+}\right) }{\partial X\left( t\right) }.
\]

The right-hand side of these equations are continuous, so we may omit the
superscript. Furthermore
\[
\frac{\partial X\left( \tau \right) }{\partial p\left( t\right) }=\left\{
X\left( t\right) ,X\left( \tau \right) \right\},\qquad \frac{\partial
X\left( \tau \right) }{\partial X\left( t\right) }=-\left\{ p\left( t\right)
,X\left( \tau \right) \right\},
\]
and computing the Poisson brackets in terms of the canonical variables
$\theta$ and $J$ we arrive at
\begin{eqnarray*}
\tilde\Phi[\xi] &=&\hbar\int dt^{\prime }\,N\left( t-t^{\prime }\right)
\left[ \frac \partial {\partial X}\left[ \left\{ X\left( t\right) ,X\left(
t^{\prime }\right) \right\} f\right] +\frac \partial {\partial p}\left[
\left\{ p\left( t\right) ,X\left( t^{\prime }\right) \right\} f\right]
\right] \\
&=&\frac \partial {\partial X}\left[ \left\{ X\left( t\right) ,\,N\left(
J,\theta \right) \right\} f\right] +\frac \partial {\partial p}\left[
\left\{ p\left( t\right) ,N\left( J,\theta \right) \right\} f\right],
\end{eqnarray*}
where we have defined $N\left( J,\theta \right) \equiv\hbar\int dt^{\prime
}\,N\left( t-t^{\prime }\right) X\left( t^{\prime }\right)$. Applying the
reduction of order procedure by substituting $X(t^\prime)$ by the Fourier
expression (\ref{fourier}) and using Eq. (\ref{noisefourier}) the function
$N(J,\theta)$ is written as,
\begin{equation}
N\left( J,\theta \right) = \sum_nX_n\left( J\right) e^{in\left[ \theta
+\Omega \left( J\right) t\right] }N_n\left( J\right),
\end{equation}
where $N_{n}(J)$ is given by Eq. (\ref{ene}). We may simplify further, by
observing that $\left\{ X,\left\{ p,N\right\} \right\} -\left\{ p,\left\{
X,N\right\} \right\} =-\left\{ N,\left\{ X,p\right\} \right\} =0$ by the
Jacobi identity and the fact that $\{X,p\}=1$, then by further manipulation
of the Poisson bracket terms we finally get
\begin{eqnarray}
\tilde\Phi[\xi]
&=&\left\{ X,\left\{ p,N\left( J,\theta \right) \right\} f\right\} -\left\{
p,\left\{ X,N\left( J,\theta \right) \right\} f\right\}   \nonumber \\
&=&\left\{ p,N\left( J,\theta \right) \right\} \left\{ X,f\right\} -\left\{
X,N\left( J,\theta \right) \right\} \left\{ p,f\right\}   \nonumber \\
&=&\left[ -\frac{\partial p}{\partial J}\frac{\partial X}{\partial \theta }+
\frac{\partial X}{\partial J}\frac{\partial p}{\partial \theta }\right]
\left[ \frac{\partial N}{\partial \theta }\frac{\partial f}{\partial J}-
\frac{\partial N}{\partial J}\frac{\partial f}{\partial \theta }\right]
=-\left\{ N,f\right\}   \label{Phifinal}
\end{eqnarray}

\subsection{Appendix B}

We first expand the environmental variables in creation and annihilation
operators
\[
\varphi _{\vec k}=\sqrt{\frac \hbar {2\omega _k}}\left[ a_ke^{-i\omega
_kt}+a_{-k}^{\dagger }e^{i\omega _kt}\right],
\]
where $\omega_k^2=k^2+M^2$. Next, recalling the definition of $\Xi $ given
in Eq. (\ref{4.7}) we may write
\[
\Xi \left( t\right) =\frac 12gM^3\int_{M^{-3}}d^3x\int \frac{d^3k}{\left(
2\pi \right) ^3}\;e^{i\vec k\cdot\vec x}\int_{p,k-p\geq M} \frac{d^3p}{
\left( 2\pi \right) ^3 }\varphi _{\vec p}\varphi _{\vec k-\vec p}
\]

To perform the integral over $x$ we introduce as in Sec. \ref{iii} the
function $\rho \left(k/M \right)= M^3\int_{M^{-3}}d^3x\;\exp(i\vec k\cdot
\vec x)$, which satisfies $\rho\left( 0\right) =1$. Then by direct
substitution we can write the correlation $\langle \Xi ( t) \Xi ( t^{\prime
})\rangle$ as
\begin{eqnarray*}
\left\langle \Xi \left( t\right) \Xi \left( t^{\prime }\right) \right\rangle
&=&\left\langle \Xi \left( t\right) \right\rangle \left\langle \Xi \left(
t^{\prime }\right) \right\rangle +\frac{\hbar ^2g^2}{16}\int \frac{ d^3k }{
\left( 2\pi \right) ^3}\frac{ d^3k^{\prime } }{\left( 2\pi \right)^3}
\rho\left({\frac{k}{M}}\right)\rho\left({\frac{k^\prime}{M}}\right) \\
&&\int_{p,k-p\geq M}\frac{d^3p}{\left( 2\pi \right) ^3}\int_{p^{\prime
},k^{\prime }-p^{\prime }\geq M}\frac{d^3p^{\prime }}{\left( 2\pi \right) ^3}
\frac{e^{-i\omega _pt}e^{i\omega _{p^{\prime }-k^{\prime }}t^{\prime
}}e^{-i\omega _{k-p}t}e^{i\omega _{p^{\prime }}t^{\prime }}}{\sqrt{\omega
_p\omega _{p-k}\omega _{p^{\prime }}\omega _{p^{\prime }-k^{\prime }}}}
\left\langle a_pa_{k-p}a_{-p^{\prime }}^{\dagger }a_{p^{\prime }-k^{\prime
}}^{\dagger }\right\rangle
\end{eqnarray*}

Now the vacuum expectation value in the last equation can be written as
\[
\left\langle a_pa_{k-p}a_{-p^{\prime }}^{\dagger }a_{p^{\prime }-k^{\prime
}}^{\dagger }\right\rangle =\left( 2\pi \right) ^6\delta^{(3)}
\left( \vec k+
\vec k^{\prime }\right) \left\{ \delta^{(3)} \left( \vec k-\vec p+\vec p
^{\prime }\right) +\delta^{(3)} \left( \vec p+\vec p^{\prime }\right)
\right\},
\]
and since in any case we deal with values of $k$ and $k^{\prime }$ much
lower than typical values of $p$ and $p^{\prime }$, we get
\begin{equation}
\left\langle \Xi \left( t\right) \Xi \left( t^{\prime }\right) \right\rangle
=\left\langle \Xi \left( t\right) \right\rangle \left\langle \Xi \left(
t^{\prime }\right) \right\rangle +\frac{\hbar ^2g^2}8\int \frac{d^3k}{\left(
2\pi \right) ^3}\rho^2\left({\frac{k}{M}}\right) \int_{p\geq M}\frac{d^3p}
{\left( 2\pi \right) ^3}\frac{e^{-2i\omega _p\left( t-t^{\prime }\right) }}{
\omega _p^2}.  \label{C.1}
\end{equation}
To obtain Eq. (\ref{qcorr}), observe that $\int d^3k\;\rho ^2(k/M)
=(2\pi)^{3} M^3. $

\subsection{Appendix C}

Let us write the integrand in the action density (\ref{4D.1}) as
\begin{equation}
\varepsilon -v(\bar x) =\frac 13 \left(\bar x-x_L\right) \left( \bar x
-x_R\right) \left(\bar x-x_{\times }\right).  \label{D.1}
\end{equation}
It can be rearranged as $\varepsilon-v(\bar x)= (1/3)(x_\times-\bar x
)[(1/4)(x_R-x_L)^2-(\bar x-(1/2)(x_R+x_L))^2]$, which suggests writting
$\bar x=(1/2)(x_R+x_L)-(1/2)(x_R-x_L)\cos (2\varphi)=
x_L+(x_R-x_L)\sin^2\varphi$, and thus
\begin{equation}
\varepsilon -v(\bar x) =\frac 1{12} \left( x_{\times }-x_L\right) \left(
x_R-x_L\right) ^2\sin ^22\varphi \;\left( 1-\kappa ^2\sin ^2\varphi \right),
\label{D.2}
\end{equation}
where $\kappa$ is defined in Eq. (\ref{4D.2}).

The equation for a classical trajectory is $\tau=\int d\bar x\;\bar p
^{-1}=\int d\bar x\; [2(\varepsilon-v(\bar x))]^{-1/2}$ and since $d\bar x
=(x_R-x_L) \sin (2\varphi) d\varphi $ we have
\begin{equation}
\tau =\sqrt{\frac 6{x_{\times }-x_L}}\int_0^\varphi \frac{d\varphi ^{\prime
} }{\sqrt{1-\kappa ^2\sin ^2\varphi ^{\prime }}}.  \label{D.3}
\end{equation}
When $\varphi $ goes from $0$ to $\pi /2$, the point describes a half orbit.
So the period $T$ is twice the result from (\ref{D.3}) when $\varphi=\pi/2$,
and one can check that when $\varepsilon \rightarrow 0$,
$T\rightarrow 2\pi$, as it should.

Using now the identities 16.1.3, 4 and 5 from Ref. \cite{AS}, we get the
desired formula (\ref{4D.3}). Observe that as $\varepsilon \rightarrow 0,$
${\rm sn}\;u\sim \sin u$, $\lambda \sim 1/2$, and $\bar x=x_L+\left(
x_R-x_L\right) {\rm sn}^2(\tau/2)\sim (1/2)\left( x_R+x_L\right)
-(1/2)\left( x_R-x_L\right) \cos \tau $ which corresponds to the harmonic
trajectory, as expected near the bottom of the potential.

\subsection{Appendix D}

Here we shall borrow an argument from Whittaker and Watson \cite{WW}. The
Fourier coefficients of the classical trajectory are
\begin{eqnarray*}
\bar x_n &=&\frac 1T\int_0^Td\tau \; e^{-in\bar \Omega \tau }\bar x\left(
\tau \right) \\
&=&x_L\delta _{n0}+\left( x_R-x_L\right) \int_0^1du\;e^{-2in\pi u}{\rm sn}
^22Ku,
\end{eqnarray*}
where we have changed to the $u$ variable, used the explicit trajectories (
\ref{4D.3}) and again redefined the integration variable in the second line.
According to 16.7 and 16.8 in Ref. \cite{AS}, we have the following
properties: ${\rm sn}\;2K\left( u+1\right) =-{\rm sn}\;2Ku$, ${\rm sn}
\;2K\left( u+2iQ\right) ={\rm sn}\;2Ku$, where $Q=K^\prime/(2K)$ with $
K^\prime\equiv K[\sqrt{1-\kappa^2}]$; and ${\rm sn}\;2Ku$ has a simple pole
at $u=iQ,$ with residue $1/( 2\kappa K)$.

To compute the Fourier coefficient (for $n\neq 0$) we consider the integral
over the anti-clockwise contour $\Gamma $ with vertices at $0,$ $1,$ $2iQ$
and $2iQ-1$. The two oblique sides cancel each other, and
\begin{eqnarray*}
\int_{-1+2iQ}^{2iQ}dz\;e^{-2in\pi z}{\rm sn}^22Kz
&=&q^{4n}\int_{-1}^0ds\;e^{-2in\pi z}{\rm sn}^22Ks \\
&=&q^{-2n}\int_0^1du \;e^{-2in\pi u}{\rm sn}^22Ku,
\end{eqnarray*}
where $q=\exp(-\pi K^\prime/K)$. Near the pole, we have (see 16.8 and 16.3.1
in Ref. \cite{AS}) ${\rm sn}\;2K\left( u+iQ\right) ={\rm sn}\left(
2Ku+iK^{\prime }\right) = (\kappa {\rm sn}\;2Ku)^{-1}= (2\kappa
Ku)^{-1}\left[ 1+O\left( u^2\right) \right], $ and from Cauchy's theorem the
integral over the contour $\Gamma$ becomes $2\pi i(4\kappa ^2K^2)^{-1}\left.
d_z\exp(-2in\pi z)\right| _{z=iQ}=n\pi ^2 \kappa ^{-2}K^{-2}q^{-n} $ which
yields Eq. (\ref{foucoef}).

\subsection{Appendix E}

Here we compute the average power exchanged between the system and the
environment. Let us go back to the Langevin equations (\ref{lange2}) with
the Gaussian source $\xi(t)$ described by Eq. (\ref{noise}). From these
equations it follows that at the phase space point $\left( X,P\right)$ there
is a dissipated power $w_d=-\Gamma P$, where $\Gamma=-\int_{-\infty}^t
dt^\prime H(t-t^\prime)X(t^\prime)$ and a noise power $w_n=\xi P$. The power
dissipated in the whole ensemble is $W_d=-\int dXdP\;f\left( X,P\right)
\Gamma P. $

Let us use the action-angle variables at $t=0$ as Lagrangian coordinates
identifying a given trajectory. Then the power dissipated is
\[
W_d=-\int_0^{J_s}dJ\;f\left( J\right) \int d\theta \;\Gamma
P=-\int_0^{J_s}dJ\;f\left( J\right) \Omega \left( J\right) {\cal D}.
\]
The average noise power in the whole ensemble is $\left\langle
W_n\right\rangle =\int_0^{J_s}dJ\;f\left( J\right) \int d\theta
\;\left\langle \xi \left( t\right) P\right\rangle $ where $P\left( \theta
,J,t\right) $ has been perturbed away from the classical value by the action
of the noise. We then get, using the Novikov trick
\[
\left\langle \xi \left( t\right) P\right\rangle =\int^tdt^{\prime }\,N\left(
t-t^{\prime }\right) \frac{\delta P\left( t\right) }{\delta j\left(
t^{\prime }\right) }=\int^tdt^{\prime }\,N\left( t-t^{\prime }\right) \frac{
\partial X\left( t^{\prime }\right) }{\partial X\left( t\right) }=-\left\{
P,N\right\}.
\]
We can simplify this by using the same arguments as in Appendix A: $\int
d\theta \{P,N\}=\int d\theta (\partial _{\theta }P)_{J}(\partial
_{J}N)_{P}=\partial _{J}\int d\theta (\partial _{\theta }P)_{J}N.$
Performing now a last integration by parts $\int d\theta \left\{ P,N\right\}
=\partial _{J}\int d\theta \;N(\partial _{\theta }P)_{J}=-\partial _{J}\int
d\theta \;P(\partial _{\theta }N)_{J}=-\partial _{J}{\cal N}.$

The total average power is thus
\[
W=W_d+W_n=\int_0^{J_s}dJ\;f\left( J\right) \;\left[ \frac{\partial {\cal N}}{
\partial J}-\Omega \left( J\right) {\cal D}\right].
\]
Integrating by parts, knowing that ${\cal N}\left( 0\right) =f\left(
J_s\right) =0$, we get $W=\int_0^{J_s}dJ\;\Omega \left( J\right) \bar\Phi
\left( J\right) $ where $\bar\Phi $ is the flux. Recalling now that $\Omega
=dE/dJ $, with our boundary conditions we have $W=E_s\bar\Phi ( E_s)
+\int_0^{J_s}dJ\;E\partial_t f. $ We can now show that $W\sim 0$ (this is
obviously true for the equilibrium solution $F_B$, when both terms in the
expression for $W$ vanish independently). Let us write the general solution
for the distribution $f$ as the following superposition $f=\int_0^\infty
dr\;e^{-rt}c_rF_r\left( j\right). $

Since the flux is linear in $f$, and the flux for $F_r$ is $K_r$, we get for
$W$,
\[
W=\int_0^\infty dr\;e^{-rt}c_r\left( E_sK_r-r\int_0^{J_s}dJ\;EF_r\right).
\]
The integral in the second term is dominated by the upper limit (since $
\sigma \sim E^{-2}$, the integral depends only logarithmically on the peak
$E^{*}$), and $\int_0^{J_s}dJ\;EF_r\sim E_s\int_0^{J_s}dJ\;F_r=E_sK_r/r,$
which makes the total averaged power $W\sim 0$ as expected.

\end{document}